\documentclass[iop]{emulateapj}
\usepackage{xcolor}
\usepackage{multirow}
\usepackage{graphicx}

\newcommand{\hii}{H\,{\sc ii} }

\slugcomment{}
\shorttitle{ALMA observations of protostellar outflows}
\shortauthors{T. Baug et al.}
\begin{document}

\title{ALMA observations reveal no preferred outflow--filament and outflow--magnetic field orientations}
\author{T. Baug\altaffilmark{1}}
\altaffiltext{}{Corresponding author: Ke Wang (kwang.astro@pku.edu.cn)}
\altaffiltext{1}{Kavli Institute for Astronomy and Astrophysics, Peking University, 5 Yiheyuan Road, Haidian District, Beijing 100871, China}
\author{Ke Wang\altaffilmark{1}}
\author{Tie Liu\altaffilmark{2}}
\altaffiltext{2}{Shanghai Astronomical Observatory, Chinese Academy of Sciences, 80 Nandan Road, Shanghai 200030, China}
\author{Mengyao Tang\altaffilmark{3}}
\altaffiltext{3}{Department of Astronomy, Yunnan University, Kunming, 650091, China}
\author{Qizhou Zhang\altaffilmark{4}}
\altaffiltext{4}{Harvard-Smithsonian Center for Astrophysics, 60 Garden Street, Cambridge, MA 02138, USA}
\author{Di Li\altaffilmark{5, 6}}
\altaffiltext{5}{CAS Key Laboratory of FAST, National Astronomical Observatories, Chinese Academy of Sciences, Beijing 100101, China}
\altaffiltext{6}{University of Chinese Academy of Sciences, Beijing 100049, China}
\author{Eswaraiah Chakali\altaffilmark{7}}
\altaffiltext{7}{National Astronomical Observatories, Chinese Academy of Science, A20 Datun Road, Chaoyang District, Beijing 100012, China}
\author{Sheng-Yuan Liu\altaffilmark{8}}
\altaffiltext{8}{Academia Sinica, Institute of Astronomy and Astrophysics, P.O. Box 23-141, Taipei 106, Taiwan}
\author{Anandmayee Tej\altaffilmark{9}}
\altaffiltext{9}{Indian Institute of Space Science and Technology, Thiruvananthapuram 695 547, Kerala, India}
\author{Paul F. Goldsmith\altaffilmark{10}}
\altaffiltext{10}{Jet Propulsion Laboratory, National Aeronautics and Space Administration, United States}
\author{Leonardo Bronfman\altaffilmark{11}}
\altaffiltext{11}{Departamento de Astronom\'{i}a, Universidad de Chile, Casilla 36-D, Santiago, Chile}
\author{Sheng-Li Qin\altaffilmark{3}}
\author{Viktor L. Toth\altaffilmark{12}}
\altaffiltext{12}{E\"{o}v\"{o}s Lor\'{a}nd University, Department of Astronomy, P\'{a}zm\'{a}ny P\'{e}ter s\'{e}t\'{a}ny 1/A, H-1117, Budapest, Hungary}
\author{Pak-Shing Li\altaffilmark{13}}
\altaffiltext{13}{University of California, Berkeley, United States}
\author{Kee-Tae Kim\altaffilmark{14}}
\altaffiltext{14}{Korea Astronomy and Space Science Institute, 776 Daedeokdae-ro, Yuseong-gu, Daejeon 34055, Republic of Korea}
%
\begin{abstract}
We present a statistical study on the orientation of outflows with respect to large-scale filaments and the magnetic
 fields. Although filaments are widely observed toward Galactic star-forming regions, the exact role of filaments
 in star formation is unclear. Studies toward low-mass star-forming regions revealed both preferred and random
 orientation of outflows respective to the filament long-axes, while outflows in massive star-forming regions
 mostly oriented perpendicular to the host filaments, and parallel to the magnetic fields at similar physical scales. 
 Here, we explore outflows in a sample of 11 protoclusters in \hii regions, a more evolved stage compared to IRDCs,
 using ALMA CO (3--2) line observations. We identify a total of 105 outflow lobes in these protoclusters. Among
 the 11 targets, 7 are embedded within parsec-scale filamentary structures detected in $^{13}$CO line and 870 $\mu m$
 continuum emissions. The angles between outflow axes and corresponding filaments ($\gamma_\mathrm{Fil}$) do not
 show any hint of preferred orientations (i.e., orthogonal or parallel as inferred in numerical models) with
 respect to the position angle of the filaments. Identified outflow lobes are also not correlated with the
 magnetic fields and Galactic plane position angles. Outflows associated with filaments aligned along
 the large-scale magnetic fields are also randomly orientated. Our study presents the first statistical results
 of outflow orientation respective to large-scale filaments and magnetic fields in evolved massive star-forming
 regions. The random distribution suggests a lack of alignment of outflows with filaments, which may be a result
 of the evolutionary stage of the clusters.
\end{abstract}

\section{Introduction}
\label{sec:intro}
{\sl Herschel} observations revealed ubiquitous filamentary structures in Galactic star-forming clouds. Filaments
 are observationally characterized as overdense elongated features of molecular clouds having an aspect ratio of more than
 $\sim$5--10 \citep{andre14}. These filaments are considered to play an important role in star formation. Dense star-forming
 cores may form within these filamentary structures \citep{andre10, arzoumanian11, lee14}. Filaments are even capable to lead
 the formation of massive stars ($m_\ast \gtrsim 8 M_\odot$) at their common junction \citep[``hub'';][]{myers09, dale11}.
 Two types of gas flows are typically expected in a star-forming filament. These could be large scale flows from the surrounding
 cloud onto the short-axis of the filament or flow of gas from the parent cloud along the long-axis of the filament
 \citep{kirk13, fernandez14}. Generally, flow of gas along the long-axis of a filament is implied by velocity gradients of
 the gas within filaments \citep{liu16a, wang18, yuan18}. Observationally such phenomena is indeed noted in several Galactic
 massive star-forming regions \citep[e.g.,][]{liu12, busquet13, lu18, baug18, yuan18}.
 
Classically, it is expected that the angular momentum of a star-forming molecular cloud is transported
 to protostars via dense cores \citep{bodenheimer95}. In a non-turbulent scenario, flow of gas along short-
 or long-axis of a filament leads to rotation of the embedded cores either parallel or perpendicular to the
 parent filament. In such condition, if the embedded protostars within the cores inherit the angular momentum
 axis, they should also follow the preferred direction of the rotation as of the cores. However, numerical studies
 showed that inflow of turbulent gas along a filament onto a core may affect the dynamics of the core and may even
 lead to fragmentation \citep[see][and references therein]{kudoh08, offner16}. It is also possible that the
 rotation axis of a protostar is independent of the natal filamentary structure. Even the angular momentum
 axes of cores were found to be distributed randomly regardless of the cloud or filamentary structures
 \citep[see e.g.,][]{goodman93, tatematsu16}. Simulations of \citet{offner16} showed that the wide-binary ($>$ 500 au) of
 slightly magnetically supercritical turbulent cores might also affect the rotation axis. Recently, \citet{lee16}
 found randomly aligned outflow axes (vis-a-vis rotation axes) of wide-binary pairs.
 
A comprehensive method to understand the influence of filaments on protostars would be identifying an explicit
 correlation between the protostellar accretion and gas flow along the filaments. But a direct detection of
 accreting gas at core scale is difficult not only because of inadequate resolution and sensitivity of current
 observational facilities, but also because of complicated gas dynamics at that scale. However, a solution to
 this problem cloud be finding correlation of the protostellar jets or bipolar outflows associated with the
 filamentary structures. A general understanding is that these bipolar outflows are launched by the rotating
 accretion disk of the protostar, and can be used to infer the orientation of the rotation axis. Also, these
 outflows are much easier to detect and identify compared to accretion disks \citep[][and references therein]{bally16}.

\begin{deluxetable*}{cccccccccc}
\tablewidth{0pt}
\tabletypesize{\scriptsize} 
\tablecaption{Details of targets\label{table1}}
\tablehead{
\colhead{Source} &    RA     &     Dec      & V$_{LSR}$    & Distance$^a$ & PA$_\mathrm{Fil}$ &  $\theta_\mathrm{B}$ & PA$_\mathrm{Gal}$ \\ 
		 &  (J2000)  &   (J2000)    & (km s$^{-1}$)&  (kpc)       &    (deg)          &       (deg)          &       (deg)}
\startdata 
IRAS 14382-6017  & 14 42 02  &  -60 30 35   &   -60.55     & 4.1$\pm$0.6  &   65   & 71$\pm$4    & 66  \\
IRAS 14498-5856  & 14 53 42  &  -59 08 56   &   -50.03     & 3.2$\pm$0.5  &   40   & 61$\pm$10   & 63  \\
IRAS 15520-5234  & 15 55 48  &  -52 43 06   &   -41.25     & 2.6$\pm$0.4  &   --   & 50$\pm$6    & 50  \\
IRAS 15596-5301  & 16 03 32  &  -53 09 28   &   -74.44     & 4.4$\pm$0.5  &   --   & 50$\pm$12   & 49  \\
IRAS 16060-5146  & 16 09 52  &  -51 54 54   &   -89.95     & 5.2$\pm$0.6  &  120   & 50$\pm$5    & 47  \\
IRAS 16071-5142  & 16 11 00  &  -51 50 21   &   -86.67     & 4.9$\pm$0.7  &   58   & 41$\pm$3    & 47  \\
IRAS 16076-5134  & 16 11 27  &  -51 41 56   &   -87.32     & 5.0$\pm$0.7  &   48   & 47$\pm$13   & 47  \\
IRAS 16272-4837  & 16 30 59  &  -48 43 53   &   -46.42     & 3.2$\pm$0.3  &   --   & 41$\pm$10   & 43  \\
IRAS 16351-4722  & 16 38 49  &  -47 28 03   &   -40.64     & 2.9$\pm$0.4  &   45   & 68$\pm$32   & 42  \\
IRAS 17204-3636  & 17 23 50  &  -36 38 58   &   -17.94     & 2.9$\pm$0.6  &   --   & 36$\pm$7    & 34  \\
IRAS 17220-3609  & 17 25 24  &  -36 12 45   &   -94.67     & 7.6$\pm$0.3  &   24   & 30$\pm$32   & 34 
\enddata 
\tablenotetext{$a$}{Distances estimated using the Kinematic Distance Calculation Tool of \citet[][http://www.treywenger.com/kd/]{wenger18} }
\tablenotetext{}{PA$_\mathrm{Fil}$ values marked with `--' where target is not associated with filaments }
\end{deluxetable*}

Recently, \citet{stephens17} explored the low-mass star-forming Perseus molecular cloud using the Submillimeter Array
 (SMA) observations and performed a statistical study on the orientation between outflows and filaments using a sample
 of fifty-seven protostars. They found a random distribution of outflow-filament orientation. Outflow orientation studies
 in massive star-forming regions are comparatively limited only to a few regions. Earlier \citet{wang11} studied
 the P1 filament in IRDC G28.34+0.06, and found outflows are orientated mostly perpendicular to the filament but
 parallel to parsec-scale magnetic fields. Recently, \citet{kong19} followed up the entire area of the same IRDC
 using the Atacama Large Millimeter/submillimeter Array (ALMA) data, and found a consistent result, i.e., the
 continuum sources which are situated on the parent filament typically have outflows directed perpendicular to the
 filament long-axis. 

Alongside the filaments, magnetic fields are also known to play a crucial role in star formation
 \citep{machida05, machida19, hull19}. In a recent study, \citet{li19} showed a possibility of perpendicular alignment of
 momentum axis in the moderately strong magnetized filaments. On the other hand, \citet{galametz18} suggested a bimodal
 distribution of outflows with respect to the magnetic field orientation. Momentum axes perpendicular to the filament are
 indeed observed in parsec-scale clumps in the massive star-forming IRDC by \citet{wang11, wang12}, and also at 0.1 pc scales
 in 7 low-mass protostellar cores by \citet{chapman13}. However, majority of these studies found a contrasting result,
 like randomly oriented outflow axes with respect to pc-scale magnetic fields \citep{targon11}. In a detailed observational
 study, \citet{zhang14} found that magnetic fields are dynamically important during the collapse of pc-scale clumps and the
 formation of sub-pc scale cores. They also reported that the role of magnetic fields is less important than gravity and
 angular momentum from the core to the disk scale by comparing core magnetic fields with the outflow axis. In a study of
 low-mass star-forming cores, \citet{hull14} also found a non-correlation between outflow axis and envelope magnetic fields.
 
Most of previous comprehensive studies are based on nearby low-mass star-forming regions with a limited number of case
 studies on massive star-forming regions \citep{wang11, wang12, kong19}. In this paper, we study outflows of 11 massive
 protoclusters \citep[1--24$\times$10$^{3}$ M$_\odot$;][]{liu16b} using ALMA data with an aim to explore the molecular
 outflows and their relations with large-scale orientation of the filaments and magnetic fields. These targets were
 carefully selected from a large sample of \hii regions \citep{liu16b}, and hence, are comparatively evolved with
 respect to the IRDC studied by \citet{wang11, wang12} and \citet{kong19}. The molecular line observations of a large sample
 of \hii regions were performed using the Atacama Submillimeter Telescope Experiment (ASTE) 10-m telescope. 
 These particular 11 targets among them showed strong blue-emission profile of HCN (4--3) which is an efficient
 tracer of infalling gas. Hence, these targets are ideal to search for the gas dynamics and the molecular outflows. Besides,
 most of these targets are embedded within large-scale filaments, while the remaining are associated with circular
 clumps. Thus, study of these targets will provide us a unique opportunity to examine the influence large-scale filamentary
 structures on the protostellar outflows in a slightly evolved massive star-forming region (i.e., \hii region) compared to
 young IRDCs \citep{wang11, kong19}, and to compare with the outflows in circular clumps. We followed up these 11 targets
 using the ALMA aiming for a comprehensive study. Detailed parameters of the target regions are presented in Table~\ref{table1}.

The distance of our target sources have been reported in \citet{faundez04}. However, we recalculated the distances using
 a renewed Galactic rotation model. The local standard of rest velocities ($v_\mathrm{lsr}$) were obtained from
 \citet{liu16b}. These distances were estimated using a python-based `Kinematic Distance Calculation Tool' of \citet{wenger18}
 which evaluates a Monte Carlo kinematic distance adopting the solar Galactocentric distance of 8.31$\pm$0.16 kpc \citep{reid14}.
 Corresponding near kinematic distances are generally agreed-well with the distance estimates reported in \citet{faundez04}. 
 The distances listed in Table~\ref{table1} are near kinematic distance to our targets. In this paper, we only present the CO
 outflows, and detailed studies on the gas dynamics and chemistry will be presented in subsequent papers. This study is organized
 as follows. Section~\ref{sec:data} describes the observations and the archival data used in the analysis. In Section~\ref{sec:result},
 we present the procedure for identifying the outflows, identification of filaments, estimation of magnetic fields' PAs
 and analysis of observed outflow parameters. Section~\ref{sec:discussion} presents a discussion of the overall scenario.
 Finally, we summarize the study in Section~\ref{sec:summary}.

\section{Data}
\label{sec:data}
\subsection{ALMA observations}
Observations of these targets were carried out from 2018 May 18 to 2018 May 20 (UTC) (ALMA Cycle 5), under the project
 2017.1.00545.S (PI: Tie Liu) using 43 12 m antennas in C43-1 configuration. The observations were obtained in four spectral
 windows in Band 7 (centering at 343.2, 345.1, 354.4, and 356.7 GHz) to cover multiple molecular lines, that are good tracers
 of infalling and outflowing gas, along with continuum. In this paper, we present CO(3--2) line observations covered in the
 345.1 GHz centered spectral band. A baseband of 1.88 GHz with spectral resolution of 1.13 MHz was used for the CO (3--2)
 observations. In these observations J1427-4206 and J1924-2914 were used as phase and bandpass calibrators while JJ1524-5903, 
 1650-5044, and J1733-3722 were observed as phase calibrators during our 3-epochs ALMA observations. We performed 
 self-calibration and cleaned the data cube using the tclean task in CASA 5.1.1. Briggs weighting with a robust
 number of 0.5 was used, and resulted a final synthesized beam size of 0$\farcs$8$\times$0$\farcs$7. The average cube
 sensitivity is 8.3 mJy beam$^{-1}$ with 1 km s$^{-1}$ wide velocity channels. We also used 0.9 mm ALMA continuum images
 and catalog in this paper to identify the driving sources of the observed outflows. Details on the identification of
 continuum sources and their analyses will be presented in a forthcoming paper.
\subsection{Molecular line data}
In order to identify the large-scale host molecular clouds and filamentary structures of our target regions, we used
 publicly available $^{13}$CO ($J = 1-0$) line maps of the Three-mm Ultimate Mopra Milky Way Survey \citep[ThrUMMS;][]{barnes15}.
 The ThrUMMS survey data has an angular resolution of 66$''$ and a velocity resolution of 0.34 km s$^{-1}$ with an rms noise
 of 0.7 K km s$^{-1}$ \citep[see Table~2 of][]{barnes15}.
\subsection{Submillimeter data}
The APEX Telescope Large Area Survey of the Galaxy \citep[ATLASGAL;][]{schuller09} imaged the inner Galactic plane 
 ($|l|\leq$60$^\circ$ and $|b|\leq$1$^\circ$.5) at  870 $\mu m$ with the Large APEX Bolometer Camera \citep[LABOCA;][]{siringo09}.
 The ATLASGAL survey data has a full width at half-maximum (FWHM) resolution of 19$\farcs$2. The ATLASGAL images were also used
 for our target regions to identify the filamentary structures.
\subsection{Dust polarization data}
Numerical studies show that the orientation of outflows depend on the direction of magnetic fields. Thus, to estimate the
 magnetic field orientation toward our target fields we obtained the {\sl Planck} \footnote{http://www.esa.int/Planck},
 353 GHz (850$\mu$m) dust continuum polarization data \citep{planck16a}. The data comprising of Stokes $I$, $Q$, and $U$ maps
 were extracted from the {\sl Planck} Public Data Release 2 \citep{planck16b} of Multiple Frequency Cutout Visualization (PR2 Full
 Mission Map with PCCS2 Catalog)\footnote{https://irsa.ipac.caltech.edu/applications/planck/}. The maps have a pixel scale of
 $\sim1\arcmin$ and beam size of $\sim5\arcmin$.

\section{Results}
\label{sec:result}
\subsection{Identification of Outflows}
For identification of outflows, we cropped the observed
 ALMA data cubes for each region into smaller cubes that only cover $\pm$200 km s$^{-1}$ centering on the systematic velocity of
 each targets. These smaller data cubes were also averaged to a resolution of 5 km s$^{-1}$ (i.e., 5 channels of original data cube)
 for enhancing the signal to noise ratio to trace the outflows more easily. We carefully examined these small data cubes looking
 for the red-blue lobes around continuum sources. It is convenient to start with the high end red- and blue-shifted velocities
 in the data cubes as these channels are least contaminated from emission from the central clouds. In addition to the bipolar
 outflows, we identified single outflows that are associated with continuum sources, and also a few outflows without having
 any association with continuum sources.

\begin{figure*}
\epsscale{1.1}
\plottwo{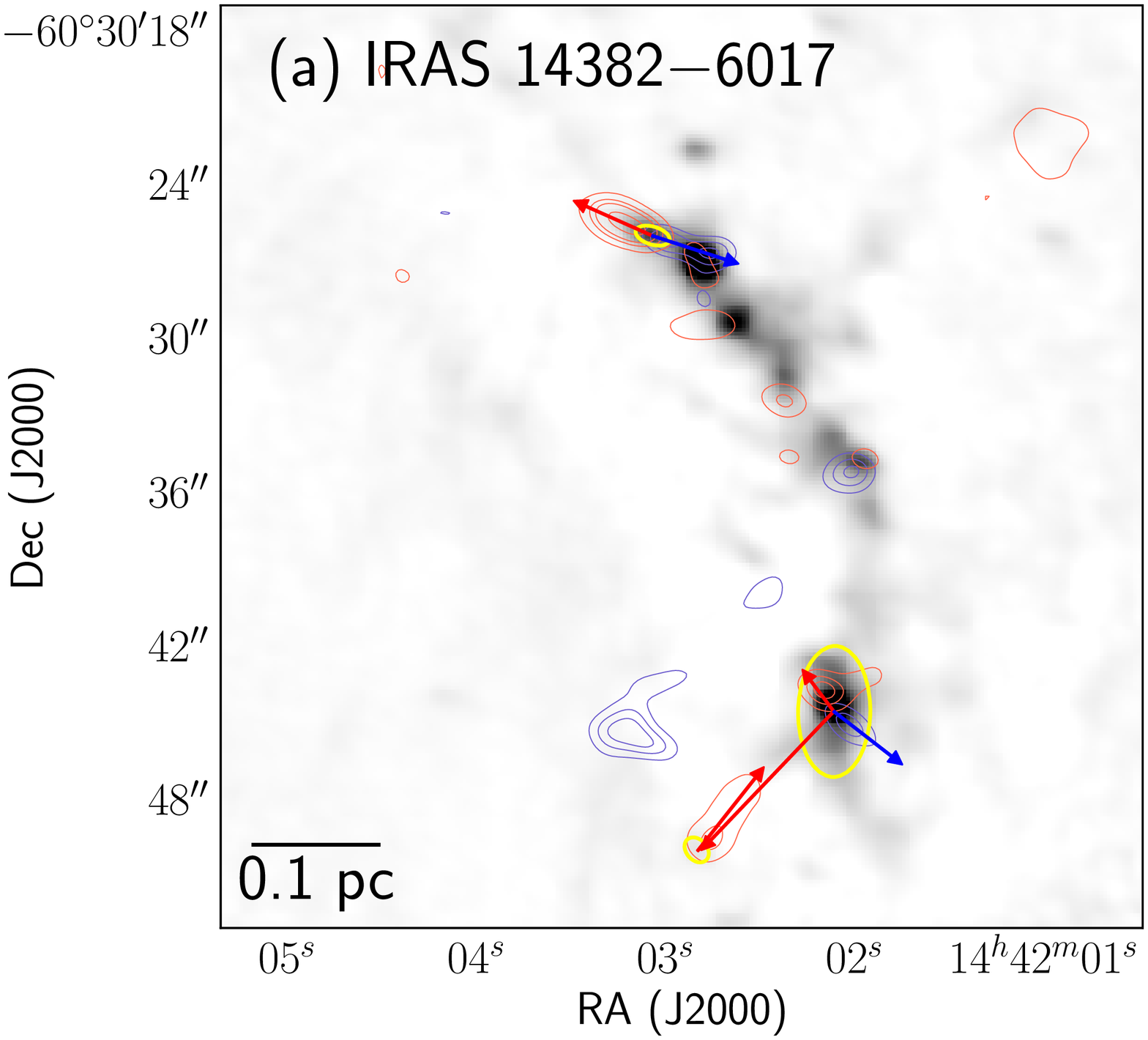}{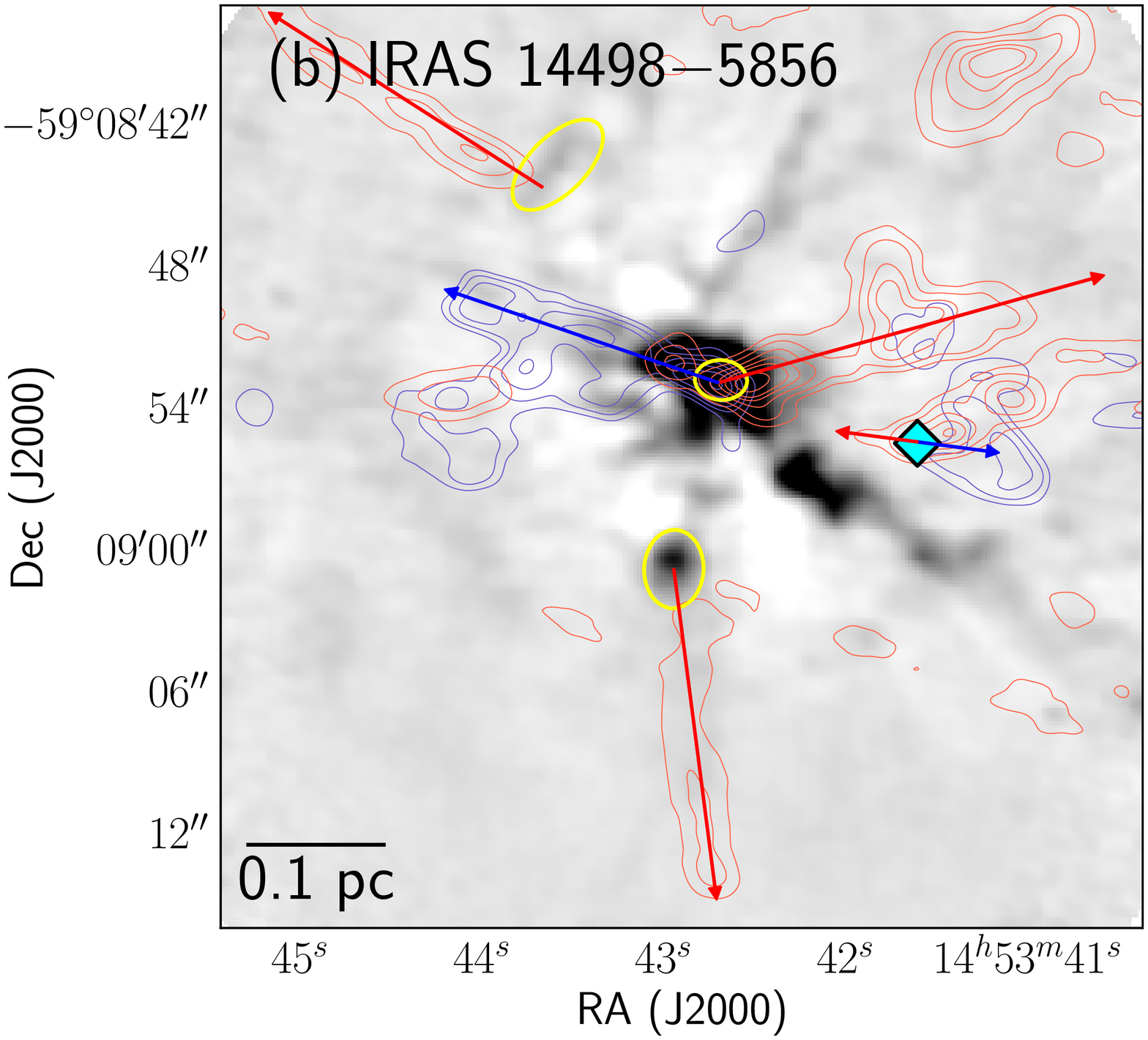}
\plottwo{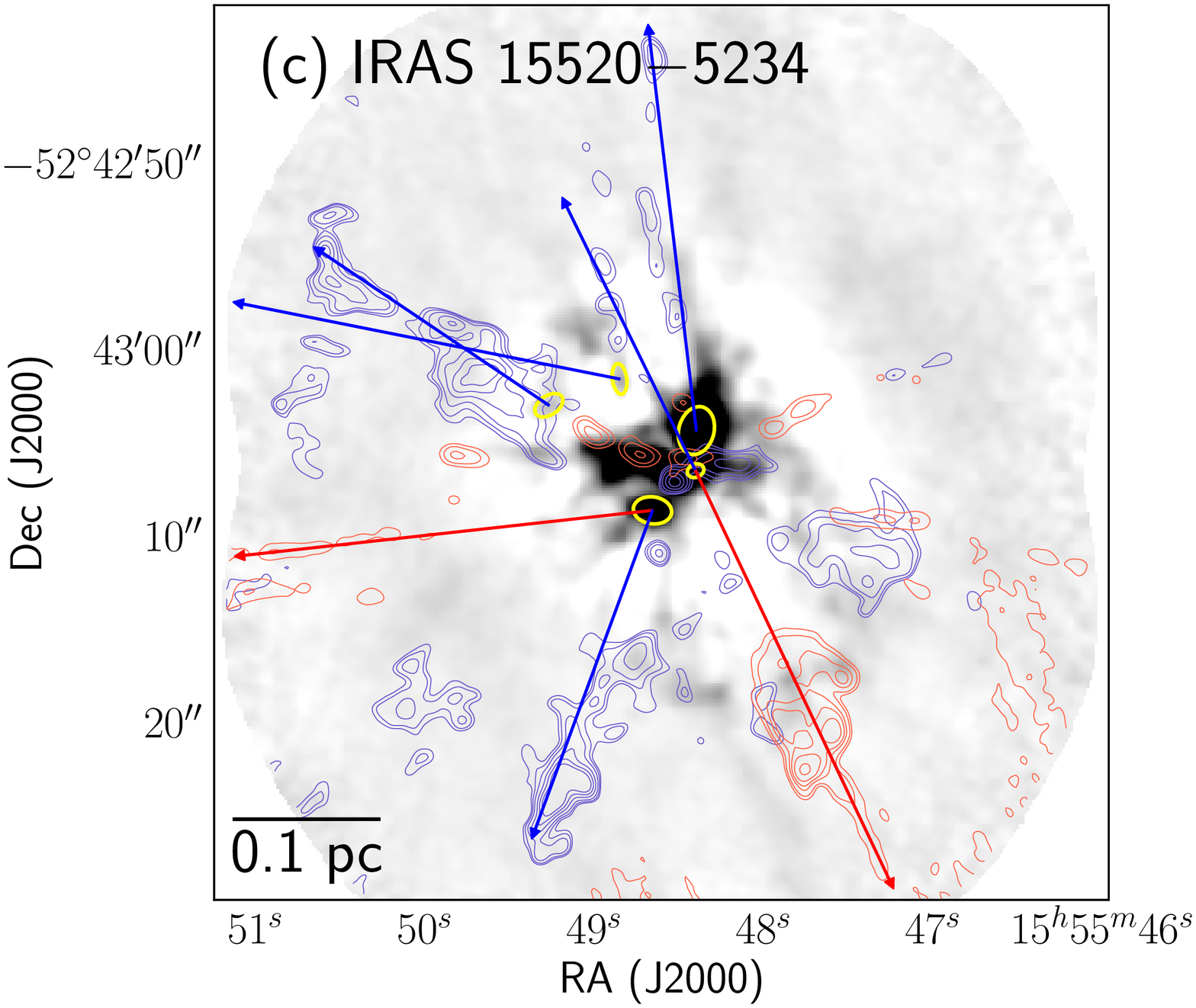}{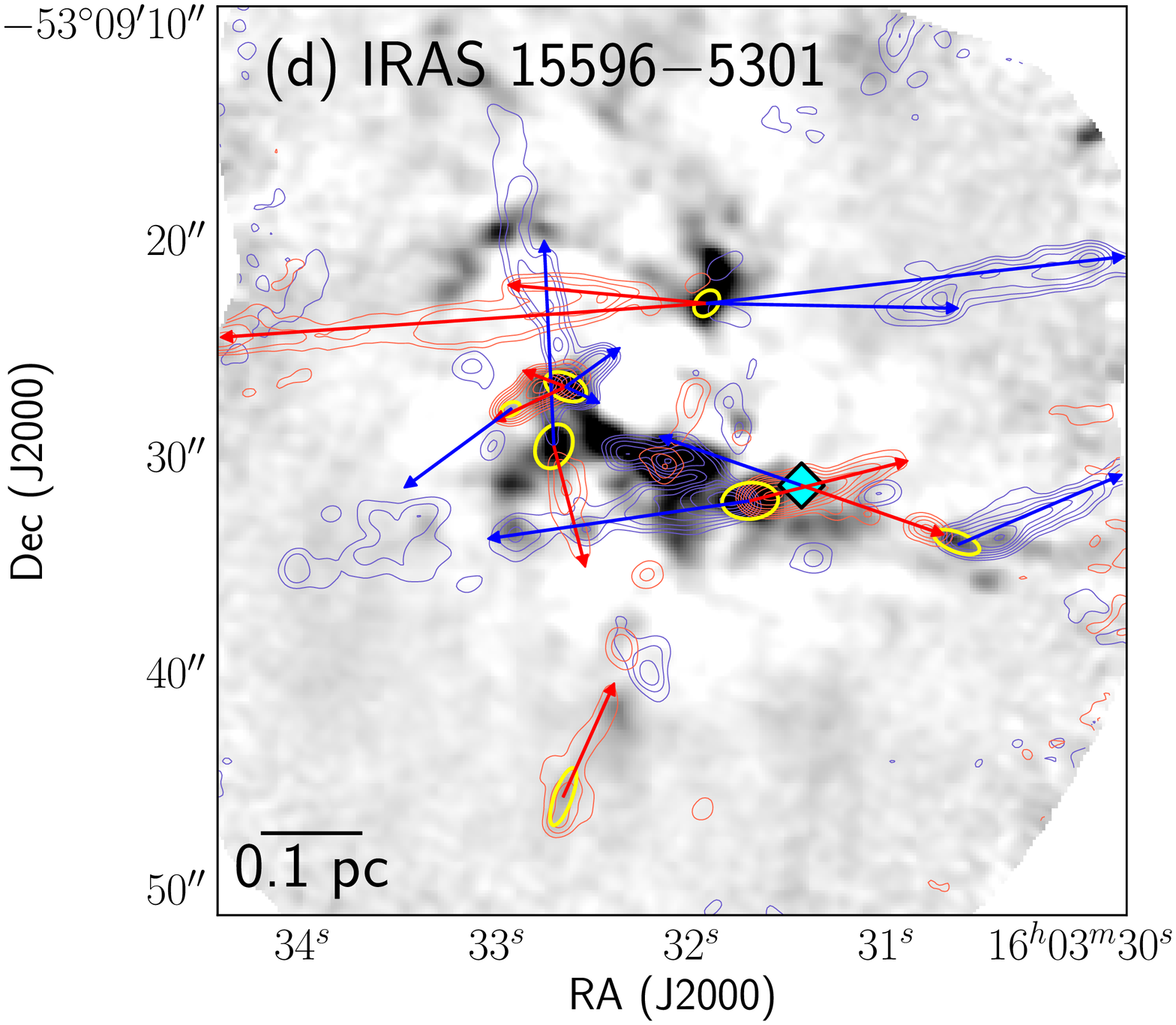}
\plottwo{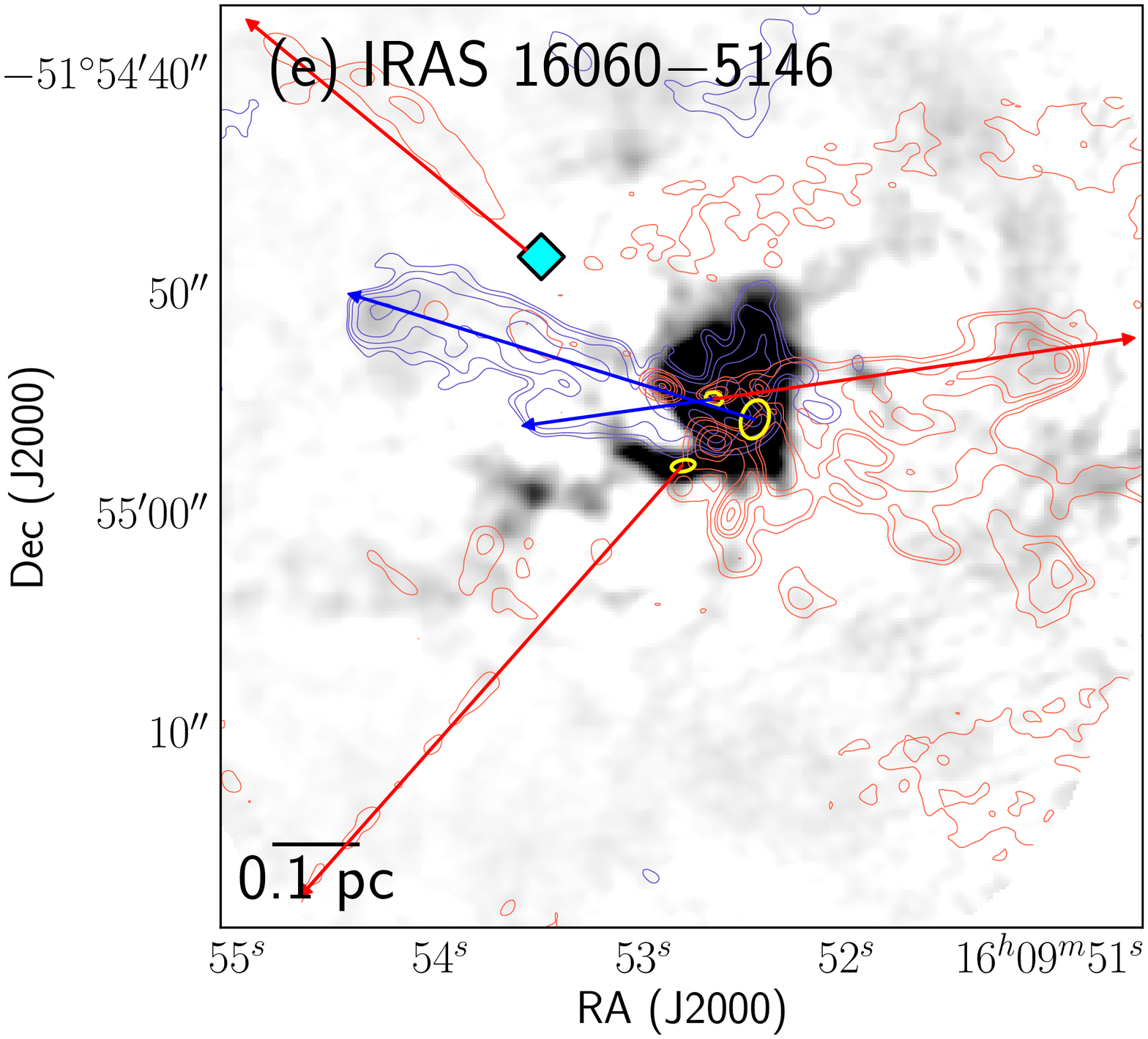}{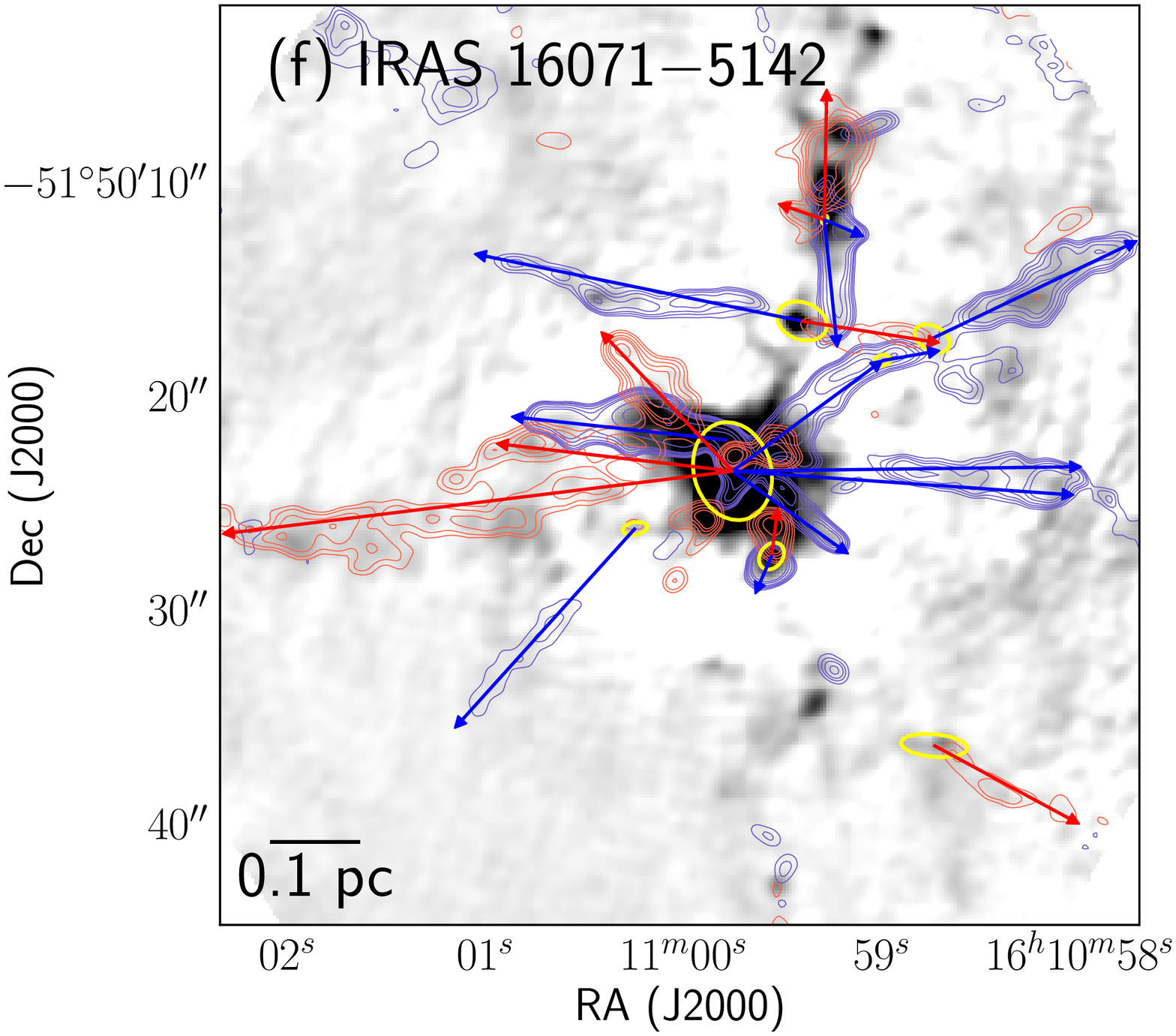}
\caption{Images of molecular outflows in the first six target fields listed in Table~\ref{table1}. Background grey-scale images are the ALMA
 0.9 mm continuum maps. The red and blue contours correspond to red and blue-shifted CO(3--2) gas integrated over carefully selected velocity
 ranges to depict the outflow lobes. The blue and red outflow lobes are also marked by blue and red arrows, respectively. The driving sources
 as shown by continuum emission are marked in yellow ellipses. Outflow lobes with cyan diamonds are those for which no continuum sources are
 identified.}
\label{fig1}
\end{figure*}
\addtocounter{figure}{-1}
\renewcommand{\thefigure}{\arabic{figure} (Cont.)}
\begin{figure*}
\epsscale{1.1}
\plottwo{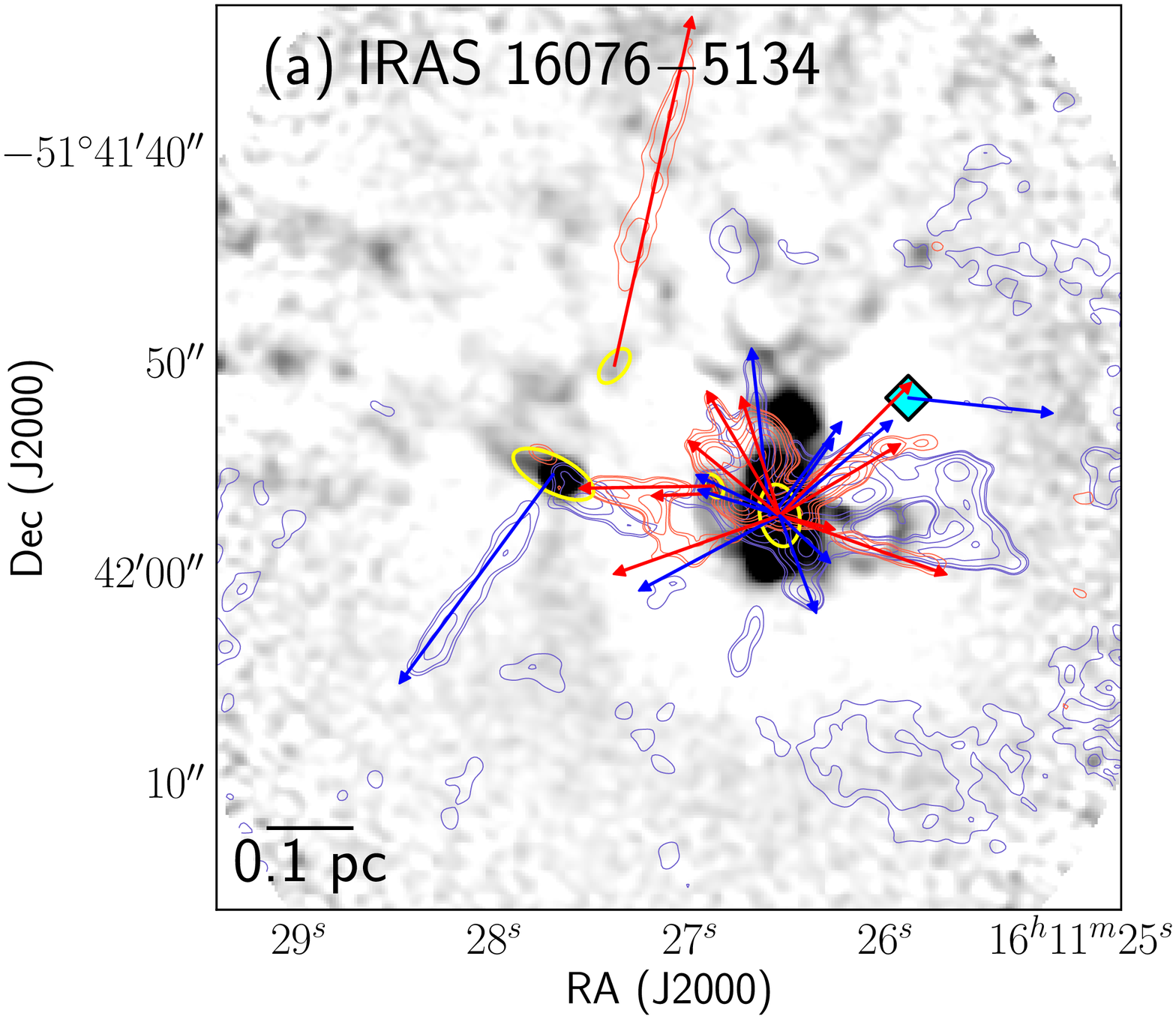}{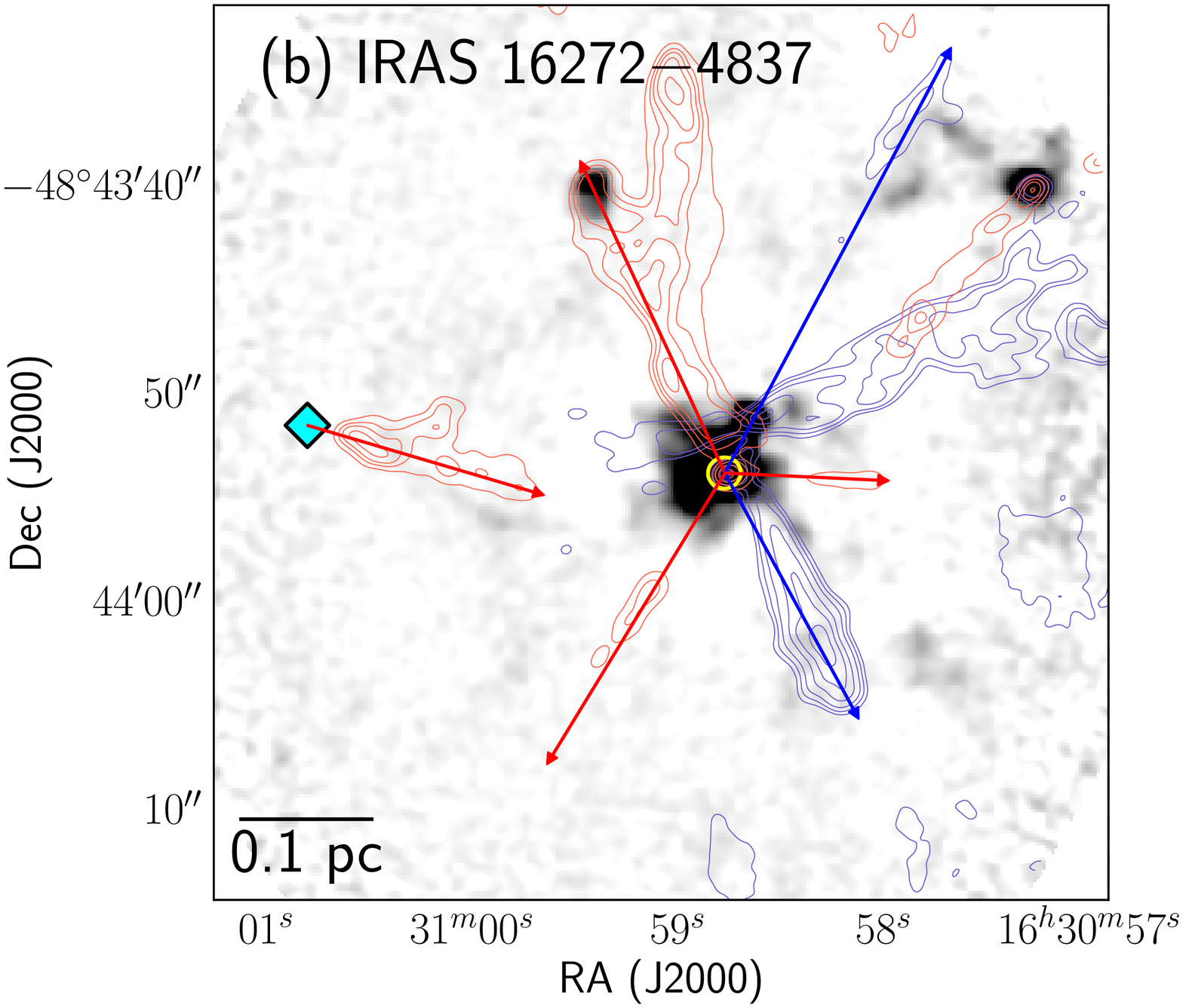}
\plottwo{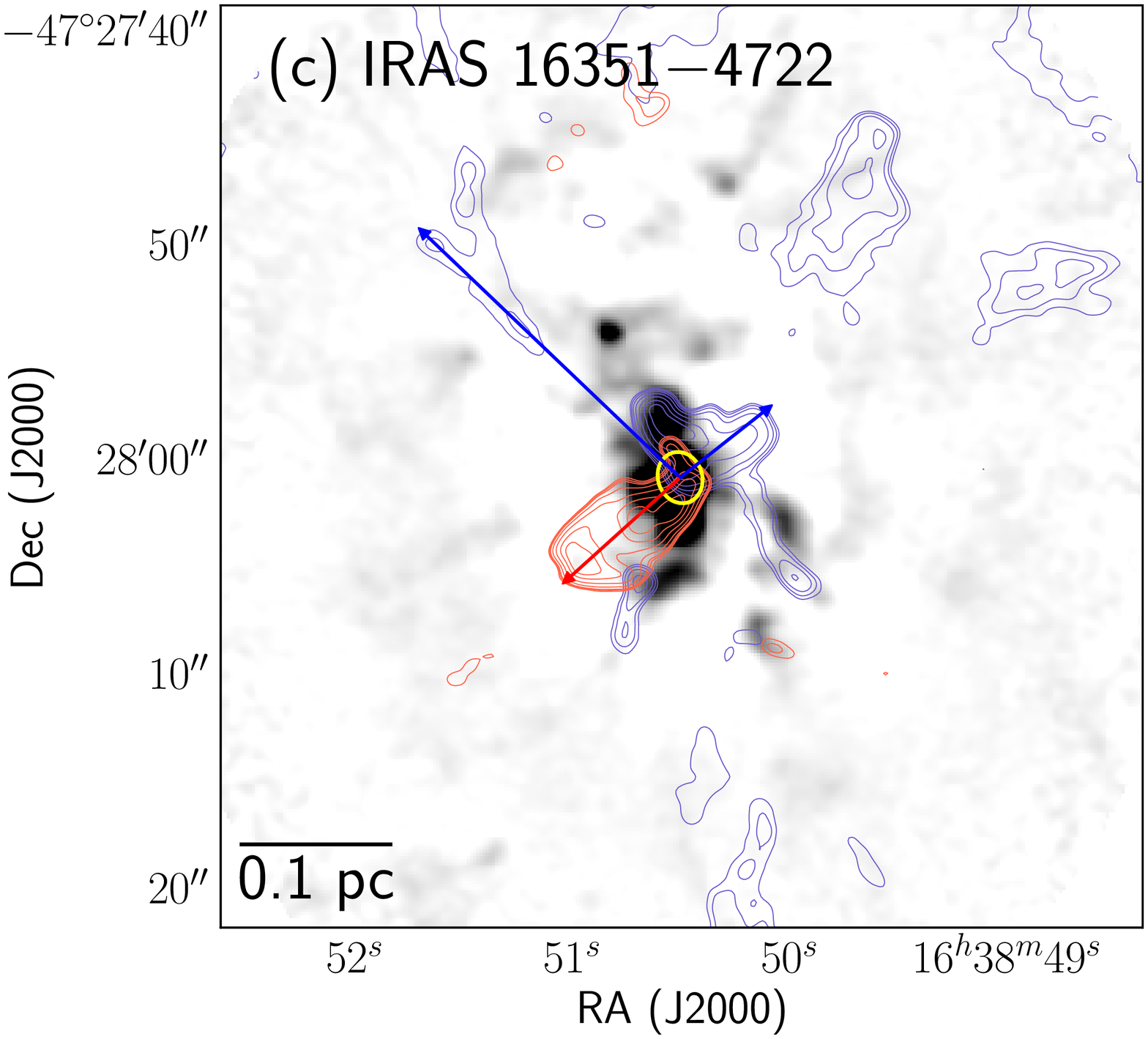}{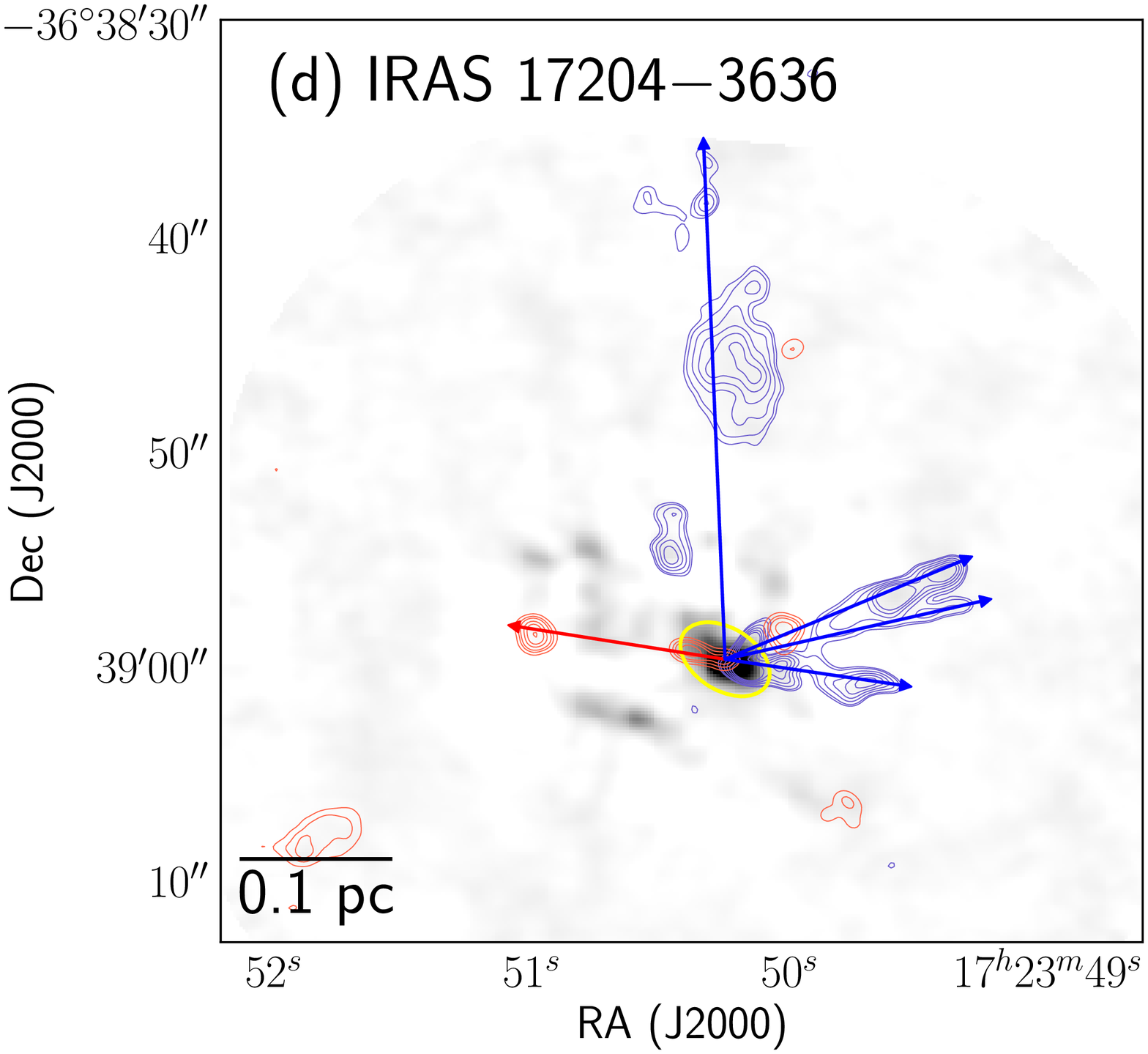}
\epsscale{0.5}
\plotone{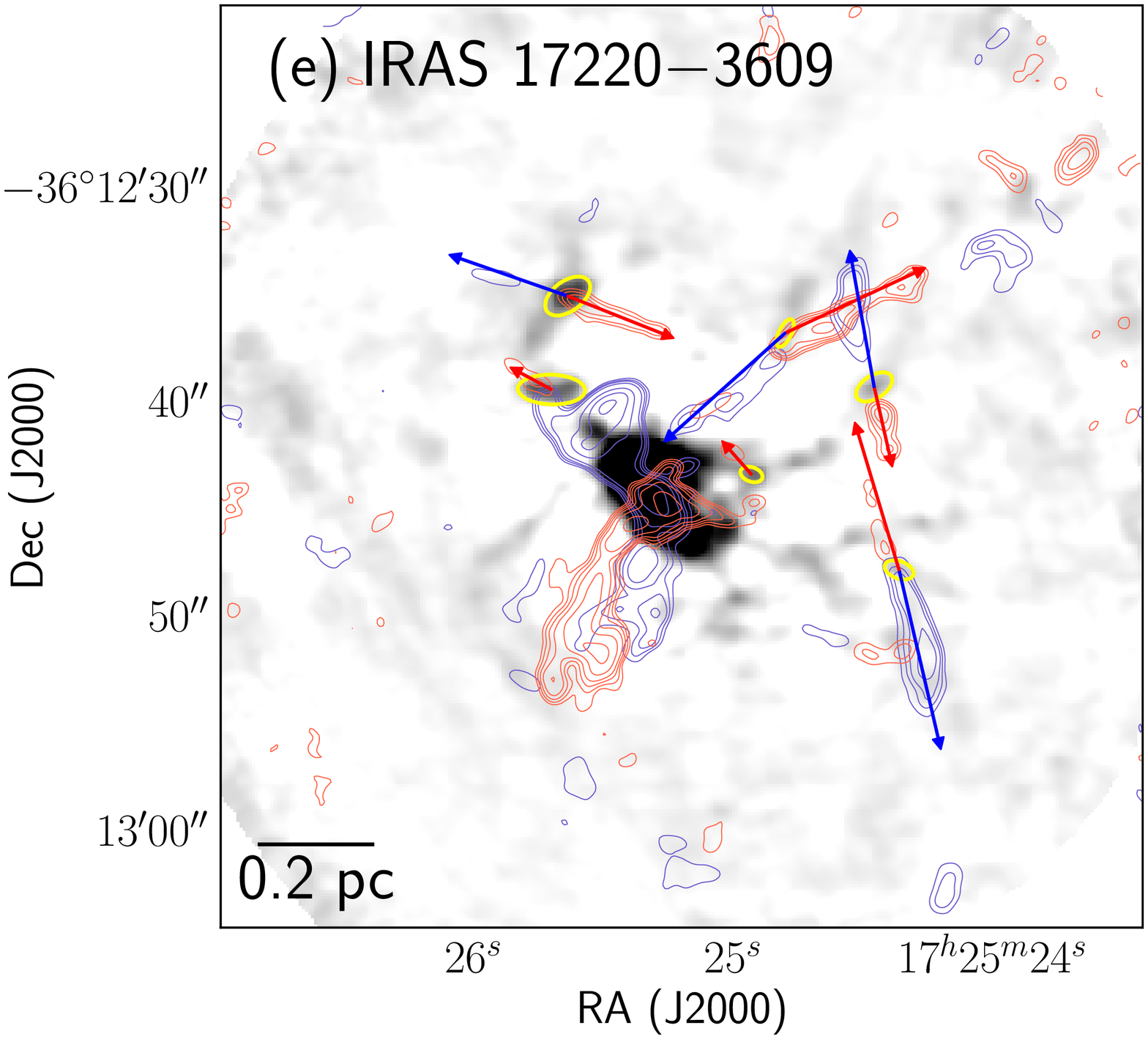}
\caption{Images of molecular outflows for the last five target fields listed in Table~\ref{table1}. The symbols are the same as in Figure~\ref{fig1}.}
\label{fig2}
\end{figure*}

After completing the preliminary identification of outflows in the averaged small data cubes, we finally inspected the original
 data cubes for the same set of outflows to obtain their final parameters. This also helped us to identify lower velocity outflows
 that were not detected in the integrated channel maps. In addition, other outflow tracers of outflows (e.g., HCN, SiO) were used to confirm
 a few confusing outflow lobes. The peak velocity and extent of each outflow were considered up to a 5$\sigma$ level where $\sigma$
 is the rms measured from a few line-free channels. Details of all identified outflows such as the coordinates of
 the continuum sources, assigned names, orientations of outflow lobes in the plane of sky, orientation of lobes with respect to underlying
 filaments, magnetic field orientation and Galactic plane (see following sections), peak velocity, and extent of each outflow are
 presented in Table~\ref{table2}. For five outflow
 lobes, no continuum sources were detected as they are possibly below our detection limit. The outflow lobes overlaid
 on the 0.9 mm ALMA continuum maps for all the regions are presented in Figure~\ref{fig1}. We finally identified a total
 of hundred and five outflow lobes. Among them 32 are bipolar and 41 are unipolar in nature.

\subsection{Identification of host clouds and filaments}
\label{SecIdenFil}
The primary aim of our study is to examine whether the orientation of outflow lobes have a dependence on large-scale filamentary
 accretion and also with the orientation of the magnetic fields or Galactic plane. Thus, we examined integrated ThrUMMS $^{13}$CO maps and the
 ATLASGAL dust continuum maps to identify the host clouds and large-scale filamentary structures. To generate the integrated
 intensity maps, velocity ranges of all the clouds were determined from the $^{13}$CO spectrum along the direction of each
 target. Priority is given to the clouds structures traced in ThrUMMS $^{13}$CO data over the ATLASGAL images as the integrated
 $^{13}$CO data for a specific velocity range suffers from least contamination from the foreground and background emission
 compared to ATLASGAL maps. However, $^{13}$CO data were not available for IRAS 17204-3636 and IRAS 17220-3609 regions. Thus,
 for these two regions, we identified clouds based on the 870 $\mu m$ ATLASGAL image.

\renewcommand{\thefigure}{\arabic{figure}}
\begin{figure*}
\epsscale{1.0}
\plotone{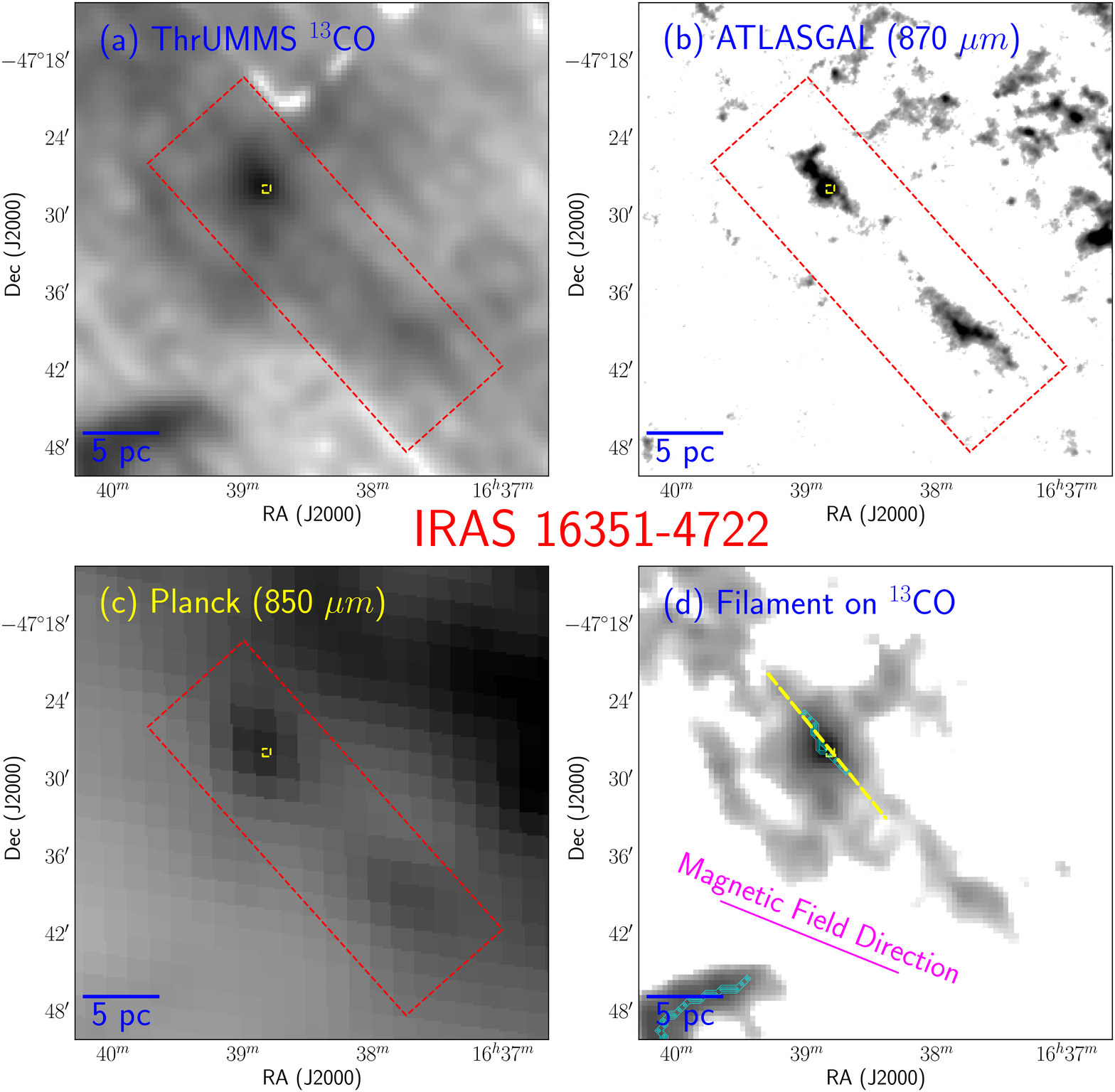}
\caption{Distribution of gas and dust in the IRAS 16351-4722 region. (a) Velocity integrated ThrUMMS $^{13}$CO map of the large
 21$'\times$21$'$ area around IRAS 16351-4722. The field of view of ALMA observations is shown by a small yellow box toward
 the center. The red rectangle shows the extent of the cloud.(b) ATLASGAL 870 $\mu m$ dust image of the same area. 
 (c) {\it Planck} 850 $\mu m$ image of the same area. (d) Filaments marked on the ThrUMMS $^{13}$CO map. The filament
 skeletons are also shown by cyan contours. Large-scale orientation of the filament is shown by an eye-fitted yellow dashed
 line. Position angle of the magnetic field is also shown by a magenta line. Magnetic field PA is estimated by averaging the
 polarization position angles of the {\it Planck} 850 $\mu m$ dust polarization map within the dashed red rectangle shown in other
 panels.}
\label{fig3}
\end{figure*}

Python-based {\sc filfinder} algorithm \citep{koch15} was applied on all the identified molecular clouds to trace the filamentary
 structures. {\sc filfinder} is capable of finding filaments even with low surface brightness as the algorithm uses an arctangent
 transform on the image. This algorithm identify all the possible filamentary structures across the input map followed by a method
 to determine their skeletons via the Medial Axis Transform. To identify the filaments, we ran the {\sc filfinder} algorithm
 with inputs like the global background thresholds and thresholds for length of skeletons. Note that the primary beam of our ALMA
 data (36$''$) is comparable or even smaller than the angular resolution of the ATLASGAL (19$\farcs$2) and ThrUMMS $^{13}$CO data
 (66$''$) making it difficult to determine the filament orientation at the scale of ALMA field of view (PA$_\mathrm{Fil}$). 
 So, we only considered the large scale PAs of filaments estimated by a visual fit over the large-scale {\sc filfinder}
 skeletons. By large-scale PAs, we imply the average PAs over at least 5 pixels of the identified skeleton (i.e., $\sim$3--4 pc at
 the distances of our targets).
 Note that a visual fit to filaments might not be as accurate as a statistical fit. However, visual fits to these large-scale 
 PA$_\mathrm{Fil}$ have typical uncertainty $\lesssim$10$\degr$. 
 We also considered elongated clumps (aspect ratio $<$ 5) as filamentary structure because such structures may also aid
 in gas-flow along a preferred direction like filaments.

\begin{figure*}
\epsscale{1.0}
\plotone{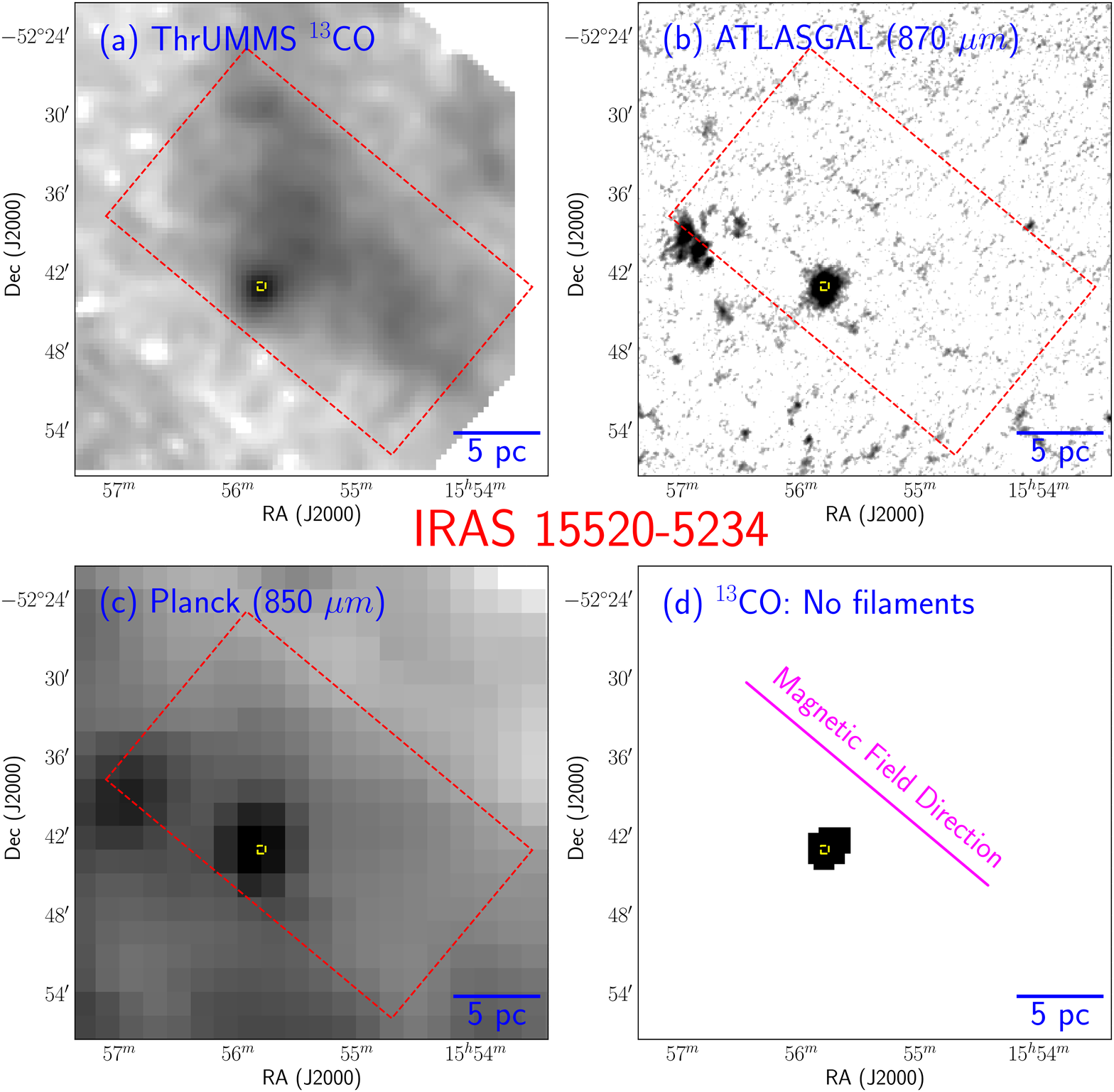}
\caption{Distribution of gas and dust in the IRAS 15520-5234 region where no filamentary structure is detected in the integrated $^{13}$CO
 and ATLASGAL 870 $\mu m$ images. Symbols are the same as in Figure~\ref{fig3}.}
\label{fig3a}
\end{figure*}

Filaments are detected toward 7 targets, and corresponding PA$_\mathrm{Fil}$ are listed in Table~\ref{table1}. The remaining
 four targets are associated with round-shaped clumps that have no preferred orientations. The distribution of integrated $^{13}$CO
 and dust emission toward two representative regions, one with filaments (IRAS 16351-4722) and another without filaments
 (IRAS 15520-5234) are shown in Figures~\ref{fig3}~and~\ref{fig3a}. The extents of the identified clouds are also marked
 in both figures. We have also shown the {\sl Planck} 850 $\mu m$ dust emission as {\sl Planck} data is used to determine
 the magnetic field PAs in our study (following section). The filament mask identified by the {\sc filfinder} algorithm,
 and a visual fit to the filament is also shown in Figure~\ref{fig3}d. Figures corresponding to the remaining targets are
 presented in Appendix~\ref{appendix}.

\subsection{Magnetic field position angle}
Our purpose of using {\sl Planck} polarization data is to infer the mean orientation of magnetic field around our targets as
 magnetic field also often aid in star formation. We estimated the mean linear polarization PAs over the cloud extent
 identified in the previous section. The conventional relation for PAs, $\theta_\mathrm{GAL} = 0.5\,\times\,arctan(U, Q)$
 (where, {\it arctan} avoids the $\pi$ ambiguity) yields
 PAs in Galactic coordinates in the range $-90\degr < \theta < +90\degr$, where $\theta_\mathrm{GAL}$~$=$~0$\degr$ pointing
 towards the Galactic North but increasing towards Galactic West. But in order to follow the IAU convention (i.e.,
 $\theta_\mathrm{GAL}$~$=$~0$\degr$ points Galactic North but increases toward Galactic East), the $\theta_\mathrm{GAL}$
 values were derived by using the relation
 
\begin{equation}
\theta_\mathrm{GAL} = 0.5\,\times\,arctan(-U,Q).
\end{equation}

The magnetic field orientations in Galactic coordinates can be obtained by adding 90$\degr$ to $\theta_\mathrm{GAL}$, i.e.,
 $\theta^{\prime}_{B}$ = 90$\degr$ + $\theta_\mathrm{GAL}$ \citep[for details see][and references therein]{planck16c, planck16d}.

Furthermore, the magnetic field orientation in celestial coordinates (FK5, J2000) is obtained using the following relation
 given in \citet{corradi98}
\begin{equation}
\psi = arctan\left[\frac{cos(l-32.9\degr)}{cos\,b~cot\,62.9\degr~-~sin\,b~sin(l-32.9\degr)}\right],
\end{equation}
where $\psi$ is the angle subtended at the position of each object by the direction of the equatorial North and the Galactic
 North. The $l$ and $b$ are the Galactic coordinates of each pixel with a polarization measurement. We then transform the
 magnetic field orientations from Galactic ($\theta^{\prime}_{B}$) to equatorial ($\theta_{B}$) coordinate system using the relation

\begin{equation}
\theta_{B} = {\theta^{\prime}_{B}} - \psi
\end{equation}

\begin{figure*}
\epsscale{1.0}
\plottwo{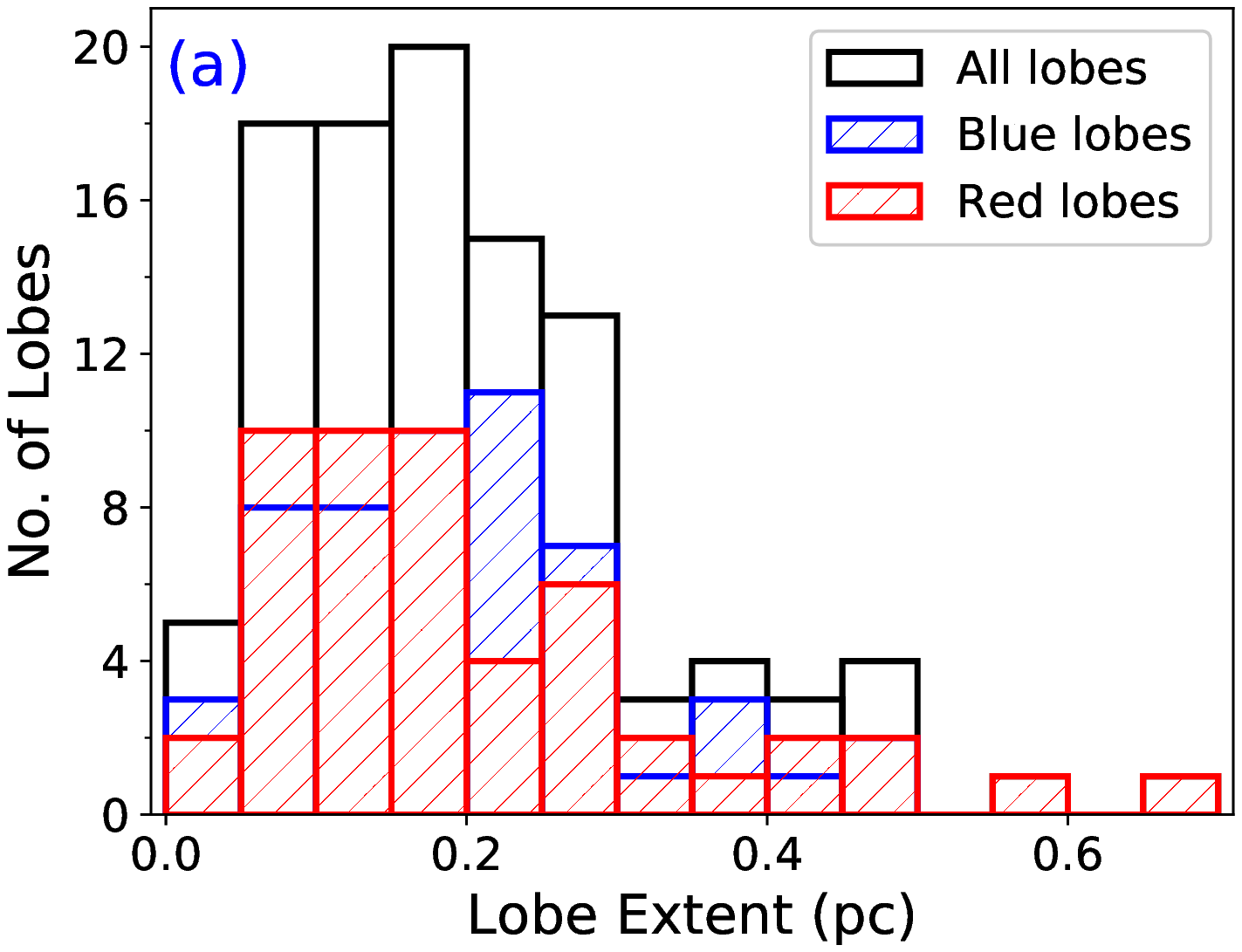}{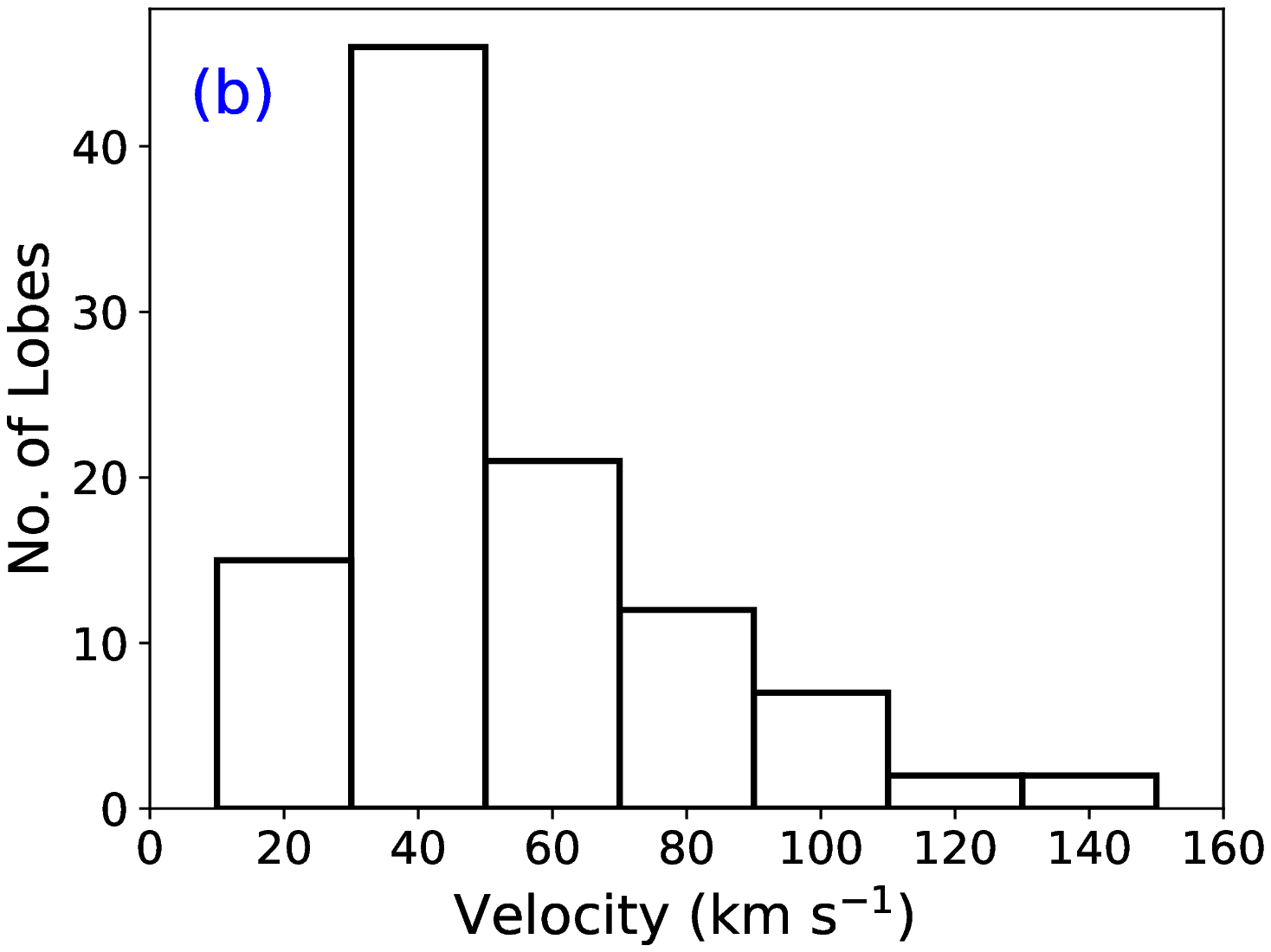}
\caption{(a) Stacked histogram of extents of all the identified outflows. Extents for blue and red-lobes are also shown
 as blue and red histograms. (b) Histogram of the peak velocities of all the identified outflows.}
\label{fig4}
\end{figure*}

Finally, we have estimated the mean magnetic field orientation using the $\theta_{B}$ vectors distributed within the area
 of the clouds identified in the previous section. The mean values of $\theta_{B}$ and the
 corresponding standard deviations are listed in Table~\ref{table1}. The mean magnetic field direction toward IRAS 14498-5856
 and IRAS 15520-5234 regions are also marked in Figures~\ref{fig3}d~and~\ref{fig3a}d.

\subsection{Statistics of Outflow extent and Velocities }
We have measured the projected plane of sky extent and maximum red-blue velocities of all identified outflow lobes. The plane
 of sky extent of outflow lobes are converted into physical scale (in pc) using the distances listed in Table~\ref{table1}.
 A histogram for all the measured outflow extents are presented in Figure~\ref{fig4}a, along with the histograms for extents
 of red and blue lobes separately. Red and blue lobes typically show a similar shape signify an unbiased identification of the outflow
 lobes. Sky projected extent of outflows lobes have a range of 0.05--0.7 pc peaking at around 0.2 pc. Similarly, we constructed
 a histogram for peak outflow velocities which is presented in Figure~\ref{fig4}b. Most of outflows have plane of sky projected 
 velocities below 50 km s$^{-1}$. However, a few outflows are seen with significantly higher velocities (up to 150 km s$^{-1}$),
 and also a few among them have extended outflow lobes more than 0.2 pc. With an average outflow extent of 0.2 pc and peak outflow
 velocity of 40 km s$^{-1}$, the typical dynamical time-scale is about 5$\times$10$^3$ yr. The average mass of the driving continuum
 sources is 15 M$_\odot$ (with T$_\mathrm{dust} \sim$ 20 K and spectral index, $\beta~\sim$2.0) and have typical outflowing mass
 of 0.5 M$_\odot$ (details are not presented in this paper). Such extended massive outflow lobes can be results of energetic
 driving protostars. Understanding of this scenario demands a detailed analysis of gas dynamics and momentum budget of the
 outflows, which will be presented in a forthcoming paper. 

 \subsection{Outflow Position Angles}
The PAs of all outflow lobes (PA$_\mathrm{lobe}$) were measured from the celestial North pole. The PAs for both blue and
 red-shifted lobes are measured independently. We are only interested in the the orientations of the lobes, and hence, alloted
 them with values in the range from -90$\degr$ to +90$\degr$ counterclockwise from the celestial North (see Table~\ref{table2}).
 Thus, the PAs in first quadrant have negative signs, and PAs in the second quadrant have positive signs. 
 
\begin{figure}
\epsscale{1.0}
\plotone{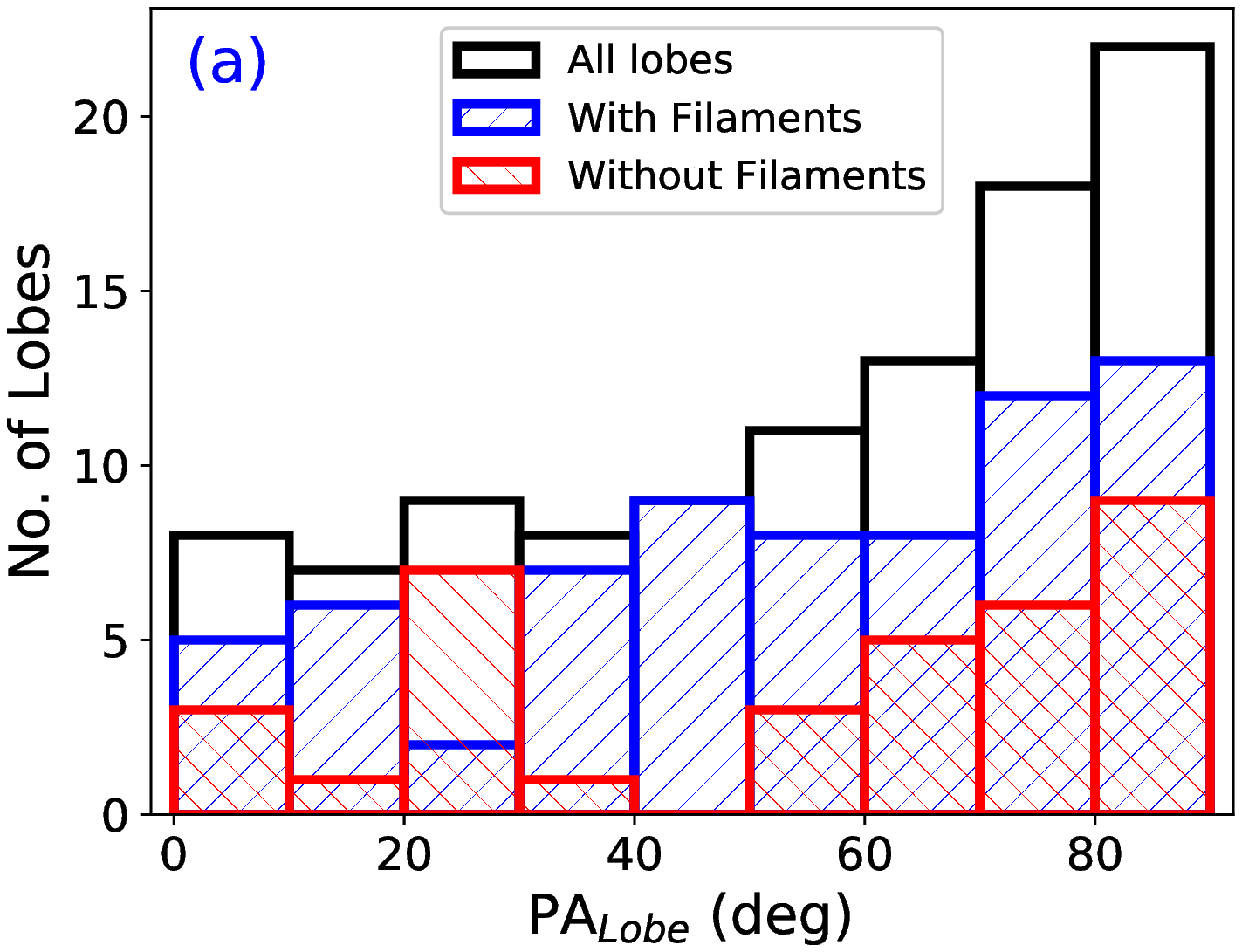}
\plotone{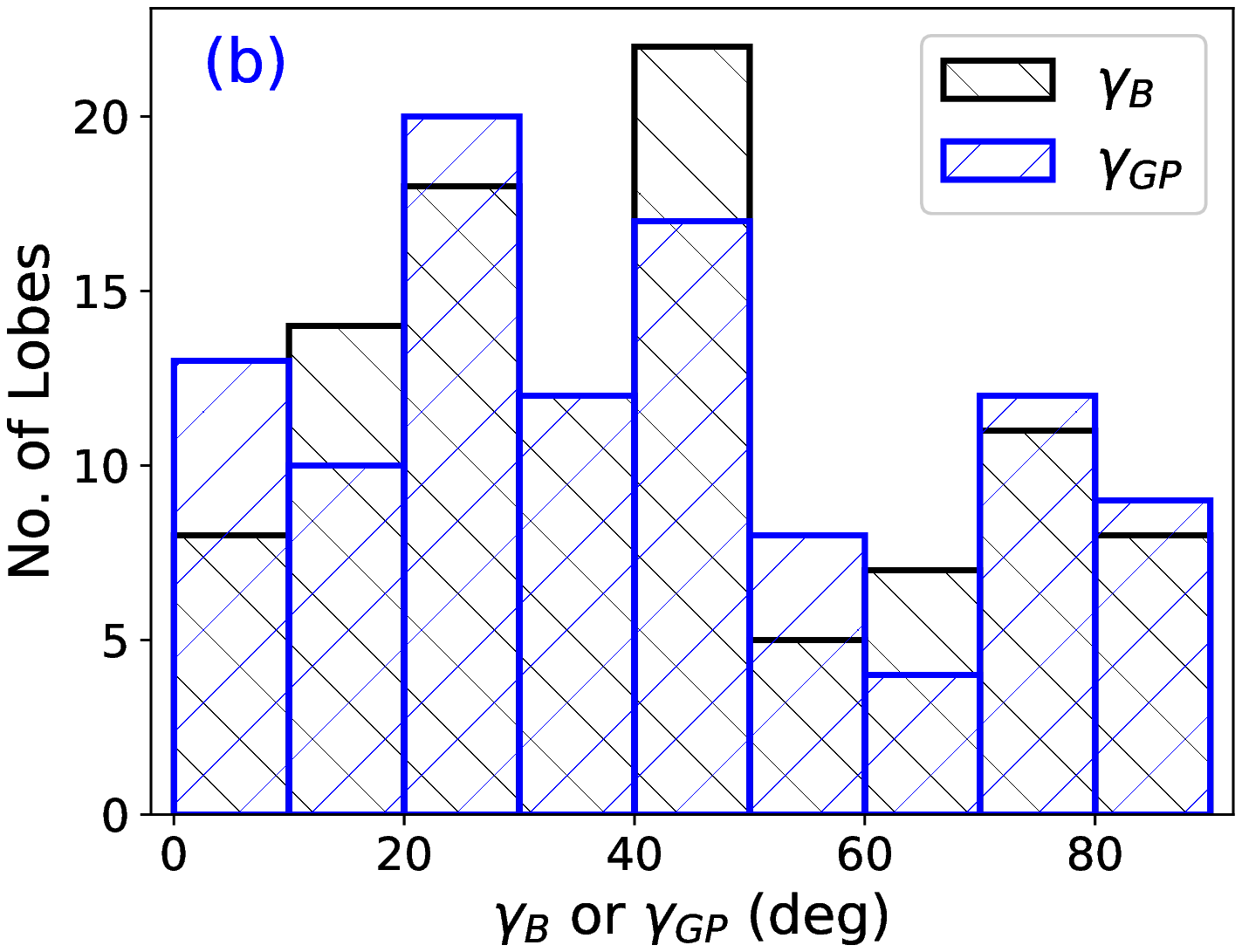}
\caption{(a) Stacked histogram of PAs of all the identified outflow lobes associated with filaments and circular clumps. (b) Stacked histograms
 of projected separations of PAs of all the lobes with respect to magnetic field orientation ($\gamma_B$) and Galactic plane ($\gamma_{GP}$).}
\label{fig5}
\end{figure}

We constructed histograms of the absolute values of the measured PA$_\mathrm{lobe}$ (ignoring the signs) considering each lobe
 as independent (Figure~\ref{fig5}). Histograms are constructed separately for the outflows associated with the filaments and
 with round-shaped/circular clumps. The outflows associated with circular clumps can be treated as a control region as they do
 not have any preferred direction of gas accretion like filaments. Although, the overall distribution of PA$_\mathrm{lobe}$ has
 a rising trend at $\sim$90$\degr$, lobes associated with filaments do not have any preferred plane of sky direction (Figure~\ref{fig5}a).
 The distribution is skewed toward PA$_\mathrm{lobe}~=$ 90$\degr$ as the outflow lobes associated with the circular
 clumps mostly oriented at PA$_\mathrm{lobe}$ in the range from 50--90$\degr$.

\subsubsection{PA$_\mathrm{lobe}$ with respect to magnetic field and Galactic plane}
As mentioned before, magnetic field plays important roles at different evolutionary phases and spatial scales of star
 formation. Hence, in this study, we also searched for any correlation of the PA$_\mathrm{lobe}$ with respect to the large-scale
 magnetic field orientation. We measured the projected plane of sky angles between PAs of large-scale magnetic field
 (see $\theta_{B}$ values in Table~\ref{table1}) and PA$_\mathrm{lobe}$ (hereafter $\gamma_{B}$). The following equation was used
 to estimate the $\gamma_{B}$ values,
 \begin{equation}
     \gamma_{B} = \mathrm{MIN}\{|\mathrm{PA}_\mathrm{lobe} - \theta_{B}|, |\mathrm{PA}_\mathrm{lobe} - \theta_{B}| - 90\degr\},
 \label{eqn}
 \end{equation}
where $\gamma_{B}$ has ranges between 0$^\circ$ and 90$^\circ$.

Note that previous studies on Galactic large-scale filaments showed most of them generally aligned with the Galactic plane \citep{wang15, wang16}
Thus, it might be important to also examine the projected plane of sky angles between PAs of Galactic plane (i.e., PA$_\mathrm{GP}$; see
Table~\ref{table1}) and PA$_\mathrm{lobe}$ (hereafter $\gamma_\mathrm{GP}$). The $\gamma_\mathrm{GP}$ values are calculated in
 similar convention as mentioned in Eq~\ref{eqn}.

Histograms for both $\gamma_{B}$ and  $\gamma_\mathrm{GP}$ are shown in Figure~\ref{fig5}b. No specific trend in the distributions is noted in 
 both the cases. This particular result indicates toward a non-correlation of the outflow axes with the large-scale magnetic field
 orientation as well as with the Galactic plane.

\subsubsection{Orientation of Outflows and Filaments}
Our primary interest in this paper is to search for the presence of any preferred angle of outflows with respect to their
 host filaments. Filamentary structures are seen in 7 regions (see Section~\ref{SecIdenFil}). These 7 targets contain
 a total of 49 outflow lobes. The PA$_\mathrm{Fil}$ values of the identified filaments are listed in Table~\ref{table1}.
 
\begin{figure}
\epsscale{1.0}
\plotone{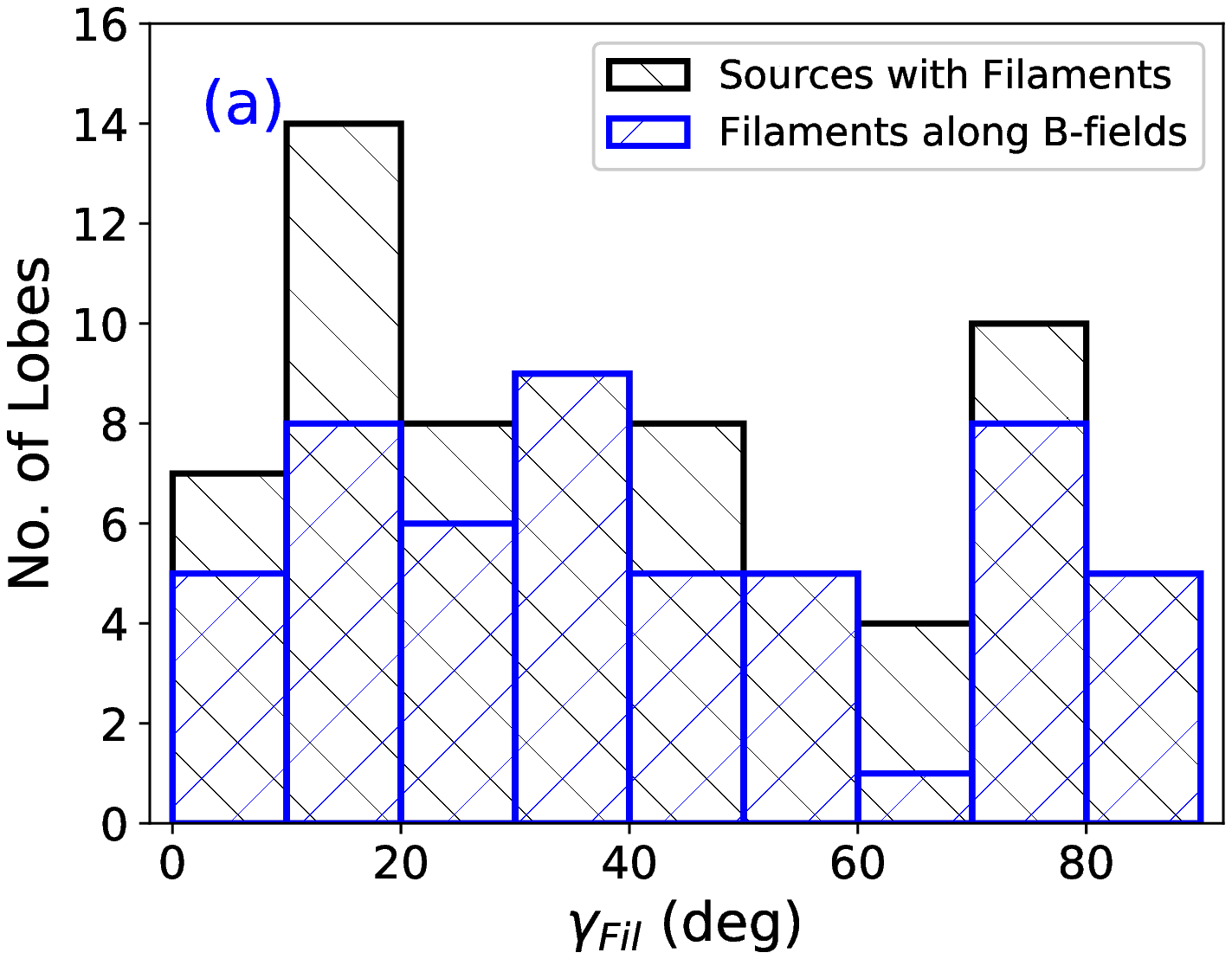}
\plotone{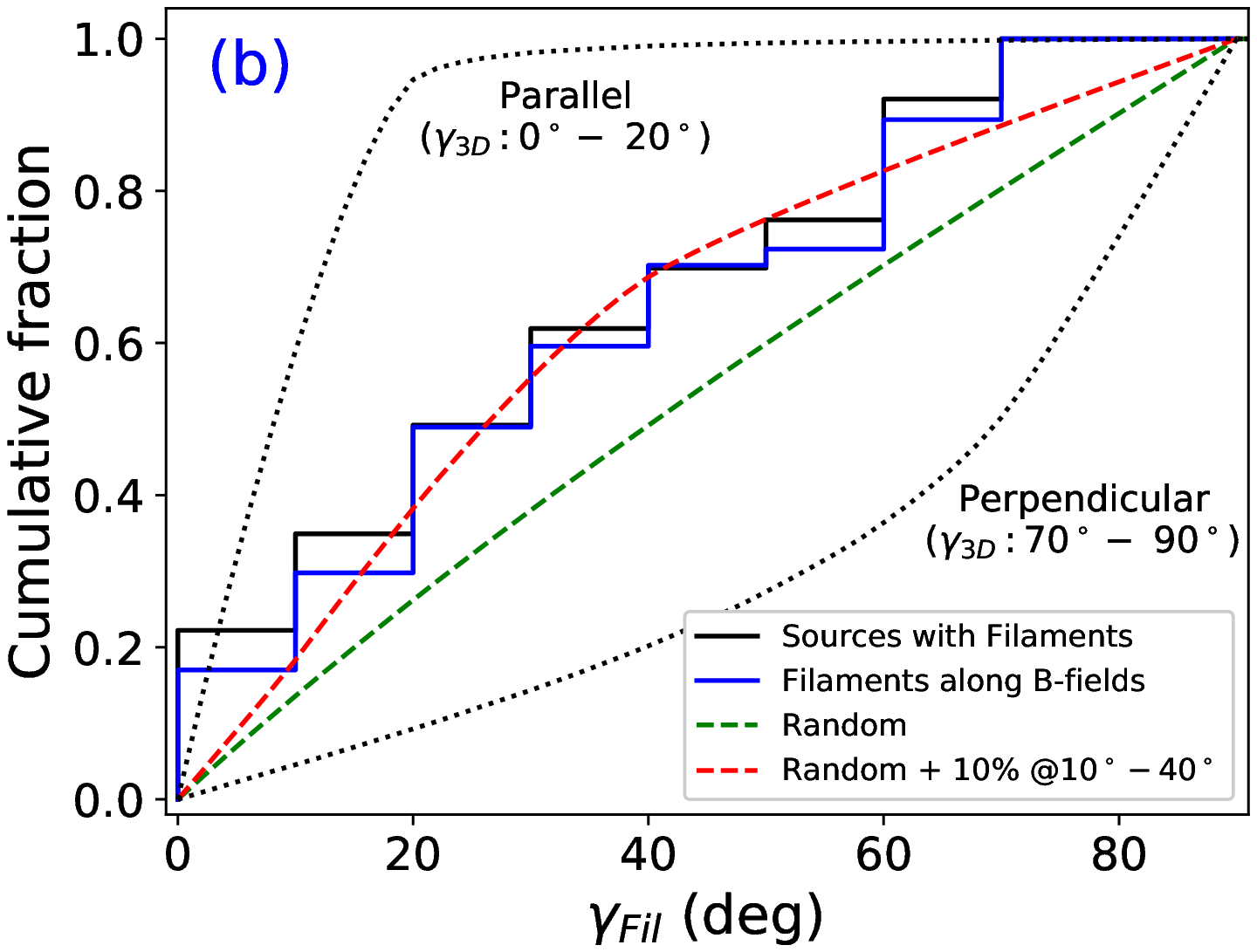}
\caption{\scriptsize (a) Histogram of $\gamma_\mathrm{Fil}$ (i.e., the angle between the PA$_\mathrm{Out}$ and PA$_\mathrm{Fil}$). 
 (b) Cumulative distribution function of $\gamma_\mathrm{Fil}$. Green line shows the CDF of randomly distributed $\gamma_\mathrm{3D}$ values.
 Two black dotted lines show parallel and perpendicular distributions of $\gamma_{3D}$ values. Red line represent the CDF 
 for the randomly distributed $\gamma_\mathrm{3D}$ values with additional 10\% sources that are oriented along 10$\degr$--40$\degr$.
 The red line better represents the observed CDF of $\gamma_\mathrm{Fil}$.}
\label{fig6}
\end{figure}

We further measured the projected PAs between PA$_\mathrm{lobe}$ and PA$_\mathrm{Fil}$ (i.e., $\gamma_\mathrm{Fil}$) for all the lobes in these 7
 regions following the same convention used in Eq~\ref{eqn}. In Figure~\ref{fig6}a, we present the histogram of all the $\gamma_{Fil}$.
 In addition, histogram of $\gamma_\mathrm{Fil}$ for those particular lobes for which the host filaments are oriented along the magnetic field (namely,
 IRAS 14382-6017, IRAS 14498-5856, IRAS 16071-5142, and IRAS 16076-5134) is also presented in Figure~\ref{fig6}a. No specific trend is
 apparent in the histograms, except both the histograms are slightly devoid of outflow lobes at $\gamma_\mathrm{Fil}\sim$60$\degr$.
 We further constructed a cumulative histogram of $\gamma_\mathrm{Fil}$ (Figure~\ref{fig6}b). It is apparent in the cumulative histogram that
 $\gamma_\mathrm{Fil}$ for all 7 regions with filaments, and also filaments aligned with magnetic fields do not show any specific trend. In fact,
 the distribution seems to be random in nature, and follow closely the random distribution curve shown in Figure~\ref{fig6}b.

Note that we measured PAs of outflow lobes and filaments on the plane of sky. The measured $\gamma_\mathrm{Fil}$ value is thus a plane
 of sky projection of the actual three-dimensional angle between outflow axes and their host filament ($\gamma_\mathrm{3D}$). Hence, the
 observed $\gamma_\mathrm{Fil}$ values might appear with a different distribution than the original distribution of $\gamma_\mathrm{3D}$ 
 \citep[see detailed discussions by][]{stephens17}. To examine the projection effect on the measured $\gamma_\mathrm{Fil}$ distribution,
 we carried out a Monte Carlo simulation. The detailed methodology can be found in \citet{stephens17}. In brief,
 we randomly generated  2$\times$10$^6$ radially outward pairs of unit vectors on the surface of a sphere. Then we calculated
 the real 3D PA between the two unit vectors ($\gamma_\mathrm{3D}$), and also their 2D PA ($\gamma_\mathrm{2D}$) assuming they are projected
 onto the y-z plane. These $\gamma_\mathrm{2D}$ values are equivalent to the observed $\gamma_\mathrm{Fil}$. Finally, we calculated the
 cumulative distribution function (CDF) of $\gamma_\mathrm{2D}$ considering three scenarios of $\gamma_\mathrm{3D}$ values:
 (1) parallel where $\gamma_\mathrm{3D}$ ranging from 0$\degr$ to 20$\degr$, (2) perpendicular where $\gamma_\mathrm{3D}$ ranging from
 70$\degr$ to 90$\degr$, and (3) random where $\gamma_\mathrm{3D}$ ranging from 0$\degr$ to 90$\degr$.  These simulated CDFs are
 also shown in Figure~\ref{fig6}b. Although, our observed CDF of $\gamma_\mathrm{Fil}$ typically follow the random distribution,
 it also deviates slightly. We thus tried a combined CDF for random and $\gamma_{3D}$ values from 10$\degr$ to 40$\degr$
 that may better represent the observed CDF of $\gamma_\mathrm{Fil}$. Accordingly, we found that the observed distribution 
 $\gamma_\mathrm{Fil}$ can be reproduced if additional 10\% sources in a random distribution have preferred $\gamma_\mathrm{3D}$ values
 from 10$\degr$ to 40$\degr$. Note that the uncertainties in the measured PAs are about $\sim$10$\degr$.
 Thus, the simulated CDF with 10\% more sources in preferred direction from 10$\degr$--40$\degr$ which better 
 represents the observed $\gamma_\mathrm{Fil}$ CDF might not be significant.
 
\begin{deluxetable*}{cclrrrrrrrrrrrrrrrr}
\tablewidth{0pt}
\tabletypesize{\scriptsize} 
\tablecaption{Outflow parameters\label{table2}}
\tablehead{
	\multicolumn{2}{c}{Continuum Source}  & Outflow &\multicolumn{2}{c}{PA$_\mathrm{lobe}$ ($^\circ$)} & \multicolumn{2}{c}{$\gamma_\mathrm{Fil}$ ($^\circ$)} & \multicolumn{2}{c}{$\gamma_{B}$ ($^\circ$)} & \multicolumn{2}{c}{$\gamma_\mathrm{GP}$ ($^\circ$)} & \multicolumn{2}{c}{$v_\mathrm{peak}$ (km s$^{-1}$)} & \multicolumn{2}{c}{Extent (pc)}\\ 
 RA (J2000)        &   Dec (J2000)    &     Name      &  Red    &  Blue  &   Red  &  Blue  &  Red   &  Blue  &  Red   &  Blue  &   Red    &  Blue  &  Red  &  Blue  }
\startdata  
14 42 02.106       & 	-60 30 44.59  &   I14382\_o1a &      37 &     52 &    -28 &    -13 &    -34 &    -19 &    -29 &    -14 &     49.5 &   46.7 & 0.041 &   0.067 \\
                   & 	              &   I14382\_o1b &     -38 &     -- &     77 &     -- &     71 &     -- &     76 &     -- &     55.6 &     -- & 0.148 &      -- \\
14 42 03.066       & 	-60 30 26.02  &   I14382\_o2  &      66 &     72 &      1 &      7 &     -5 &      1 &      0 &      6 &     70.4 &   72.7 & 0.067 &   0.070 \\
14 42 02.833       & 	-60 30 49.99  &   I14382\_o3  &     -38 &     -- &     77 &     -- &     71 &     -- &     76 &     -- &     31.3 &     -- & 0.083 &      -- \\
14 53 42.681       & 	-59 08 52.88  &   I14498\_o1a &     -74 &     -- &     73 &     -- &     45 &     -- &     43 &     -- &    119.0 &     -- & 0.263 &      -- \\
                   & 	              &   I14498\_o1b &      -- &     71 &     -- &     38 &     -- &     10 &     -- &      8 &       -- &   65.6 &    -- &   0.190 \\
$\ast\ast$         & 	$\ast\ast$    &   I14498\_o2  &      82 &     83 &     49 &     50 &     21 &     22 &     19 &     20 &     65.3 &   69.1 & 0.054 &   0.054 \\
14 53 43.579       & 	-59 08 43.78  &   I14498\_o3  &      57 &     -- &     24 &     -- &     -4 &     -- &     -6 &     -- &     19.3 &     -- & 0.213 &      -- \\
14 53 42.941       & 	-59 09 00.87  &   I14498\_o4  &       6 &     -- &    -27 &     -- &    -55 &     -- &    -57 &     -- &     25.4 &     -- & 0.151 &      -- \\
15 55 48.398       & 	-52 43 06.53  &   I15520\_o1  &      25 &     26 &     -- &     -- &    -25 &    -24 &    -25 &    -24 &     28.1 &   32.6 & 0.313 &   0.206 \\
15 55 48.654       & 	-52 43 08.66  &   I15520\_o2a &     -84 &     -- &     -- &     -- &     46 &     -- &     46 &     -- &     42.0 &     -- & 0.284 &      -- \\
                   &                  &   I15520\_o2b &      -- &    -20 &     -- &     -- &     -- &    -70 &     -- &    -70 &       -- &   32.6 &    -- &   0.236 \\
15 55 48.393       & 	-52 43 04.38  &   I15520\_o3  &      -- &      7 &     -- &     -- &     -- &    -43 &     -- &    -43 &       -- &   23.9 &    -- &   0.276 \\
15 55 48.848       & 	-52 43 01.61  &   I15520\_o4  &      -- &     79 &     -- &     -- &     -- &     29 &     -- &     29 &       -- &   61.2 &    -- &   0.266 \\
15 55 49.265       & 	-52 43 03.02  &   I15520\_o5  &      -- &     56 &     -- &     -- &     -- &      6 &     -- &      6 &       -- &   68.1 &    -- &   0.191 \\
16 03 31.921       & 	-53 09 22.96  &   I15596\_o1a &     -86 &    -84 &     -- &     -- &     44 &     46 &     45 &     47 &     41.7 &   48.4 & 0.480 &   0.415 \\
                   & 	              &   I15596\_o1b &      85 &     89 &     -- &     -- &     35 &     39 &     36 &     40 &     33.1 &   54.5 & 0.196 &   0.248 \\
16 03 32.646       & 	-53 09 26.82  &   I15596\_o2a &     -65 &    -54 &     -- &     -- &     65 &     76 &     66 &     77 &     88.5 &   51.0 & 0.079 &   0.066 \\
                   & 	              &   I15596\_o2b &      69 &     64 &     -- &     -- &     19 &     14 &     20 &     15 &     72.9 &   84.9 & 0.045 &   0.037 \\
16 03 32.656       & 	-53 09 45.76  &   I15596\_o3  &     -24 &     -- &     -- &     -- &    -74 &     -- &    -73 &     -- &     55.6 &     -- & 0.123 &      -- \\
16 03 32.705       & 	-53 09 29.57  &   I15596\_o4  &      15 &      3 &     -- &     -- &    -35 &    -47 &    -34 &    -46 &     49.5 &   42.4 & 0.123 &   0.203 \\
16 03 31.697       & 	-53 09 32.09  &   I15596\_o5  &     -76 &    -82 &     -- &     -- &     54 &     48 &     55 &     49 &    104.1 &   95.3 & 0.160 &   0.261 \\
$\ast\ast$         & 	$\ast\ast$    &   I15596\_o6  &      71 &     71 &     -- &     -- &     21 &     21 &     22 &     22 &     72.1 &   51.0 & 0.148 &   0.148 \\
16 03 32.927       & 	-53 09 27.85  &   I15596\_o7  &      -- &    -53 &     -- &     -- &     -- &     77 &     -- &     78 &       -- &   33.7 &    -- &   0.130 \\
16 03 30.635       & 	-53 09 33.99  &   I15596\_o8  &      -- &    -68 &     -- &     -- &     -- &     62 &     -- &     63 &       -- &   82.3 &    -- &   0.175 \\
16 09 52.650       & 	-51 54 54.86  &   I16060\_o1  &     -82 &    -82 &    -22 &    -22 &     48 &     48 &     51 &     51 &     40.1 &   34.5 & 0.488 &   0.223 \\
16 09 52.450       & 	-51 54 55.79  &   I16060\_o2  &      -- &     73 &     -- &    -47 &     -- &     23 &     -- &     26 &       -- &   50.1 &    -- &   0.488 \\
$\ast\ast$         & 	$\ast\ast$    &   I16060\_o3  &      50 &     -- &    -70 &     -- &      0 &     -- &      3 &     -- &     39.2 &     -- & 0.416 &      -- \\
16 09 52.803       & 	-51 54 57.90  &   I16060\_o4  &     -42 &     -- &     18 &     -- &     88 &     -- &    -89 &     -- &     26.2 &     -- & 0.662 &      -- \\
16 10 59.750       & 	-51 50 23.54  &   I16071\_o1a &     -83 &    -89 &     44 &     38 &     56 &     50 &     50 &     44 &     70.2 &   56.3 & 0.571 &   0.387 \\
                   & 	              &   I16071\_o1b &      83 &     86 &     30 &     33 &     42 &     45 &     36 &     39 &     31.2 &   35.5 & 0.264 &   0.380 \\
                   & 	              &   I16071\_o1c &      43 &     54 &    -10 &      1 &      2 &     13 &     -4 &      7 &     52.9 &   39.9 & 0.213 &   0.158 \\
                   & 	              &   I16071\_o1d &      -- &     84 &     -- &     31 &     -- &     43 &     -- &     37 &       -- &  142.2 &    -- &   0.240 \\
                   & 	              &   I16071\_o1e &      -- &    -75 &     -- &     52 &     -- &     64 &     -- &     58 &       -- &   90.1 &    -- &   0.465 \\
16 10 59.553       & 	-51 50 27.51  &   I16071\_o2  &      -7 &    -23 &    -60 &    -76 &    -48 &    -64 &    -54 &    -70 &    148.3 &  109.2 & 0.054 &   0.045 \\
16 10 59.400       & 	-51 50 16.52  &   I16071\_o3  &      78 &     81 &     25 &     28 &     37 &     40 &     31 &     34 &     31.2 &   28.6 & 0.371 &   0.154 \\
16 11 00.242       & 	-51 50 26.22  &   I16071\_o4  &      -- &    -42 &     -- &     85 &     -- &    -83 &     -- &    -89 &       -- &   19.1 &    -- &   0.299 \\
16 10 58.732       & 	-51 50 36.37  &   I16071\_o5  &      62 &     -- &      9 &     -- &     21 &     -- &     15 &     -- &     31.2 &     -- & 0.183 &      -- \\
16 10 59.286       & 	-51 50 11.78  &   I16071\_o6  &      71 &     67 &     18 &     14 &     30 &     26 &     24 &     20 &     33.8 &   94.5 & 0.054 &   0.047 \\
16 10 59.286       & 	-51 50 11.78  &   I16071\_o7  &      -1 &      6 &    -54 &    -47 &    -42 &    -35 &    -48 &    -41 &     87.6 &   65.9 & 0.145 &   0.142 \\
16 11 26.540       & 	-51 41 57.32  &   I16076\_o1a &      51 &     47 &     14 &     10 &      4 &      0 &      4 &      0 &     85.2 &   41.4 & 0.136 &   0.080 \\
 	           & 	              &   I16076\_o1b &      18 &     20 &    -19 &    -17 &    -29 &    -27 &    -29 &    -27 &    111.2 &   27.5 & 0.144 &   0.120 \\
	           & 	              &   I16076\_o1c &     -59 &     -- &     84 &     -- &     74 &     -- &     74 &     -- &     68.7 &     -- & 0.162 &      -- \\
	           & 	              &   I16076\_o1d &      76 &     64 &     39 &     27 &     29 &     17 &     29 &     17 &     80.9 &   87.3 & 0.066 &   0.107 \\
	           & 	              &   I16076\_o1e &     -62 &     -- &     81 &     -- &     71 &     -- &     71 &     -- &     35.8 &     -- & 0.184 &      -- \\
	           & 	              &   I16076\_o1f &     -45 &     -- &    -82 &     -- &     88 &     -- &     88 &     -- &     35.8 &     -- & 0.216 &      -- \\
	           & 	              &   I16076\_o1g &      71 &     72 &     34 &     35 &     24 &     25 &     24 &     25 &     35.8 &   63.1 & 0.205 &   0.098 \\
	           & 	              &   I16076\_o1h &     -50 &    -71 &    -87 &     72 &     83 &     62 &     83 &     62 &     32.3 &   31.0 & 0.170 &   0.203 \\
	           & 	              &   I16076\_o1i &      -- &      6 &     -- &    -31 &     -- &    -41 &     -- &    -41 &       -- &   56.1 &    -- &   0.194 \\
	           & 	              &   I16076\_o1j &      -- &     30 &     -- &     -7 &     -- &    -17 &     -- &    -17 &       -- &   32.7 &    -- &   0.166 \\
	           & 	              &   I16076\_o1k &      -- &    -35 &     -- &    -72 &     -- &    -82 &     -- &    -82 &       -- &   52.7 &    -- &   0.108 \\
	           & 	              &   I16076\_o1l &      -- &    -37 &     -- &    -74 &     -- &    -84 &     -- &    -84 &       -- &   35.3 &    -- &   0.128 \\
16 11 27.384       & 	-51 41 50.21  &   I16076\_o2  &     -12 &     -- &    -49 &     -- &    -59 &     -- &    -59 &     -- &     35.8 &     -- & 0.412 &      -- \\
16 11 27.697       & 	-51 41 55.36  &   I16076\_o3  &      -- &    -36 &     -- &    -73 &     -- &    -83 &     -- &    -83 &       -- &   36.2 &    -- &   0.299 \\
16 11 26.876       & 	-51 41 55.92  &   I16076\_o4  &     -89 &     -- &     54 &     -- &     44 &     -- &     44 &     -- &     36.6 &     -- & 0.155 &      -- \\
16 11 26.876       & 	-51 41 55.92  &   I16076\_o5  &      -- &    -88 &     -- &     55 &     -- &     45 &     -- &     45 &       -- &   37.9 &    -- &   0.076 \\
$\ast\ast$         & 	$\ast\ast$    &   I16076\_o6  &      -- &     84 &     -- &     47 &     -- &     37 &     -- &     37 &       -- &   23.2 &    -- &   0.167 \\
16 30 58.770       & 	-48 43 53.89  &   I16272\_o1a &      25 &     29 &     -- &     -- &    -16 &    -12 &    -18 &    -14 &     68.0 &   41.2 & 0.256 &   0.209 \\
                   & 	              &   I16272\_o1b &     -31 &    -28 &     -- &     -- &    -72 &    -69 &    -74 &    -71 &     22.9 &   20.4 & 0.254 &   0.359 \\
                   &                  &   I16272\_o1c &      87 &     -- &     -- &     -- &     46 &     -- &     44 &     -- &     31.6 &     -- & 0.122 &      -- \\
$\ast\ast$         & 	$\ast\ast$    &   I16272\_o2 &       74 &     -- &     -- &     -- &     33 &     -- &     31 &     -- &     47.2 &     -- & 0.184 &      -- \\
16 38 50.501       &    -47 28 00.91  &   I16351\_o1a &      58 &     41 &     19 &      2 &    -10 &    -27 &     16 &     -1 &     92.0 &   21.6 & 0.090 &   0.092 \\
                   &                  &   I16351\_o1b &      -- &     46 &     -- &      7 &     -- &    -22 &     -- &      4 &       -- &   35.4 &    -- &   0.237 \\
17 23 50.249       &    -36 38 59.66  &   I17204\_o1a &      81 &     82 &     -- &     -- &     45 &     46 &     47 &     48 &     20.2 &   40.5 & 0.144 &   0.123 \\
                   &                  &   I17204\_o1b &      -- &    -77 &     -- &     -- &     -- &     67 &     -- &     69 &       -- &   32.7 &    -- &   0.179 \\
                   &                  &   I17204\_o1c &      -- &    -67 &     -- &     -- &     -- &     77 &     -- &     79 &       -- &   35.3 &    -- &   0.175 \\
                   &                  &   I17204\_o1d &      -- &      2 &     -- &     -- &     -- &    -34 &     -- &    -32 &       -- &   24.9 &    -- &   0.342 \\
17 25 25.635       &    -36 12 35.12  &   I17220\_o1  &      68 &     71 &     68 &     71 &     38 &     41 &     34 &     37 &     28.4 &   32.3 & 0.196 &   0.215 \\
17 25 24.796       &    -36 12 36.85  &   I17220\_o2  &     -65 &    -48 &    -65 &    -48 &     85 &    -78 &     81 &    -82 &     97.8 &   49.6 & 0.265 &   0.280 \\
17 25 24.357       &    -36 12 47.89  &   I17220\_o3  &      13 &     17 &     13 &     17 &    -17 &    -13 &    -21 &    -17 &     35.4 &   55.7 & 0.316 &   0.264 \\
17 25 24.453       &    -36 12 39.36  &   I17220\_o4  &      13 &     10 &     13 &     10 &    -17 &    -20 &    -21 &    -24 &     33.6 &   52.2 & 0.140 &   0.238 \\
17 25 24.926       &    -36 12 43.44  &   I17220\_o5  &      41 &     -- &     41 &     -- &     11 &     -- &      7 &     -- &     30.2 &     -- & 0.076 &      -- \\
17 25 25.697       &    -36 12 39.48  &   I17220\_o6  &      62 &     -- &     62 &     -- &     32 &     -- &     28 &     -- &     34.5 &     -- & 0.080 &      -- 
\enddata
\tablenotetext{$\ast\ast$}{No continuum sources are detected for these outflows.}
\end{deluxetable*}

\section{Discussion}
\label{sec:discussion}
The projected momentum axes of outflow lobes in our studied protoclusters do not show any preferred direction with respect to 
 the observed filaments. The $\gamma_\mathrm{Fil}$ are in fact distributed in a random fashion (see Figure~\ref{fig6}b). For further
 confirmation of the randomness, we performed a statistical Kolmogorov--Smirnov (K--S) test on the observed CDF of $\gamma_\mathrm{Fil}$
 with the simulated CDFs for random and random$+$preferred(10\% in 10$\degr$--40$\degr$) $\gamma_\mathrm{3D}$ values. 
 Both the tests produce $p-$values more than 0.8 which imply that we cannot reject the null hypothesis at a level of 80\% or lower.
 This particular test indicates that the distribution is likely to be random in nature (with at least 80\% confidence).
 In addition, our identified outflow lobes do not show any preferred orientation with respect to the large-scale magnetic field
 as well as with the Galactic plane. 
 
Random orientation of outflow lobes on the plane of sky for every origin indeed refers that the distribution of $\gamma_\mathrm{Fil}$
 is really random in nature. Note that we identified filaments with a visual fit to the {\sc filfinder} skeletons, and the uncertainties
 in the large-scale PA$_\mathrm{Fil}$ are typically $\lesssim$10$\degr$. Thus, we also constructed cumulative distribution of
 $\gamma_\mathrm{Fil}$
 values by adding a random number in range of $\pm$10$\degr$. The cumulative distribution does not show any significant change
 compared to the original curve. Two such examples are shown in Figure~\ref{figA10} in Appendix~\ref{appendixB}. Thus, a robust
 identification of filaments may only slightly improve the statistics but not the overall finding of this study.  

Filamentary structures in molecular clouds may develop through several physical processes, and accretion through filaments
 also vary depending on the presence of the magnetic fields \citep[see][for more detailed discussion]{stephens17}. Observational
 studies of \citet{hacar13} and \citet{pineda11} suggest that filaments may fragment into smaller substructures, which may
 significantly affect the initial conditions for protostellar accretion and collapse. These prolate shaped substructures
 are generally aligned along the filaments \citep{pineda11, hacar13}, but this is not the case always \citep{pineda15}. Formation
 of protostars within these arbitrarily distributed smaller substructures may thus lead to a random direction of outflows.
 
Magnetic fields are known to play a crucial role in both low- and high-mass star formation \citep{hull19} but only at tens
 of pc to sub-pc scales \citep{li13,zhang14,santos16}. The role of the magnetic field becomes less important compared to
 gravity and angular momentum at the core to disk scale \citep[0.01 pc;][]{zhang14}. In a study of low-mass star-forming
 cores, \citet{hull14} found no correlation between outflow axis and envelope magnetic fields. Simulation of \citet{li15}
 showed that the local small scale (i.e., clump/core scale) magnetic field and filament orientations can be substantially
 different from the large scale orientations. The deviation in orientation also depends on the gas density of the cores
 and filaments. Even if the local magnetic field aligned with the large scale magnetic field, the turbulence within cores
 may lead to misalignment of outflow axes and magnetic field \citep{gray18}. With {\sl Planck} polarization
 data, we could only estimate the large-scale (a few pc) magnetic field which was compared with the sub-pc scale driving
 sources. The internal dynamics and magnetic field at core-scale could be highly different from that measured in the 
 large-scale. This could be a possible reason of non-correlation of $\theta_{B}$ with PA$_\mathrm{lobe}$.
 
Earlier, \citet{davis09} and \citet{stephens17} also found a random alignment of molecular outflows (i.e., momentum axes) with
 respect to the filament/core directions in nearby Galactic star-forming regions. In contrast, \citet{anathpindika08}, \citet{wang11},
 and \citet{kong19} found that the momentum axes of outflows are preferentially oriented perpendicular to the filaments. Numerical
 simulations suggest both scenarios are possible depending upon the initial condition of the host cloud that formed the filaments,
 on their surrounding environment, and the presence of magnetic fields. In fact, momentum axes of outflows may vary significantly
 depending on how exactly the accretion occurs to the central protostars. For example, on the one hand, \citet{clarke17} showed
 that accretion onto a filament occurring from a turbulent environment may produce vorticity which have angular momentum parallel
 to the axis of the filament. 
 
On the other hand, magnetic fields also play a crucial role in star formation \citep{machida05, machida19, hull19}.
 Simulations show that orientation of outflows with respect to their parent cores (thus, the filaments) could depend strongly
 on the relative strengths of the magnetic field, turbulence, and rotation \citep{machida05, offner16, lee17}. It is important
 to note that although the feedback from star formation is energetically important \citep{arce11} and is capable of sustaining
 turbulence even in a low mass star forming region \citep{lih15}, the dynamic flow seems to de-couple from filament in the
 protostellar accretion phase based on the results here. Recently, \citet{li19} suggested for a possibility of perpendicular
 alignment of momentum axis in the moderately strong magnetized filaments. While some observations revealed magnetic field
 lines to be perpendicular to the orientation of the filament \citep[see e.g.,][]{matthews00, santos16}, \citet{galametz18}
 showed a bimodal distribution outflows with respect to envelope-scale magnetic field in a few protostars. Earlier,
 \citet{wang12} observed a filament perpendicular and corresponding outflow lobes parallel to the magnetic fields.
 
Most of the previous studies are based on the nearby star-forming clouds. Observational evidences for both preferred and
 random orientation of outflow lobes are found in the studied regions. The only distant massive star-forming region with 
 comprehensive outflow orientation study is IRDC G28.34+0.06 \citep{wang11, wang12, kong19}, and outflows in this region orient
 perpendicular to the underlying filaments. However, our study with 11 massive protocluster show a contrasting result.
 Note that the regions presented in this paper have already appeared with \hii regions that are characteristics of newborn massive
 stars. These regions are at a later evolutionary stage compared to the infrared dark clouds \citep[for example studied
 by][]{wang11, kong19}. With the evolutionary sequence, the outflow power declines and the primary winds tend to dominate over
 the outflows with the evolution of prestellar cores \citep{bally16}. In addition, multiplicity is also a common phenomena
 in the dense massive star-forming environment. Interaction with companions may significantly affect the protostar's spin
 \citep{offner16, lee17}, and hence, may lead to a randomly oriented outflow lobes.

Another possibility of random distribution of outflow lobes could be the operating of multiple mechanisms in the same 
 molecular cloud. This is because while some simulations suggest momentum axis parallel to the filament axis under
 certain conditions, others are suggestive of perpendicular momentum axis depending upon a different physical condition
 of the surrounding environment. Thus, in a combined
 environment and physical condition, one may ideally expect to see a random orientation of the outflow axes with respect to
 the filament axis. \citet{stephens17} tried to disentangle the observed outflows assuming that they are not purely random in 
 nature, and found a hint for momentum axes tending to align perpendicular to the filament axis. Although not very significant,
 our analysis shows that $\gamma_\mathrm{Fil}\sim$60$^\circ$ is devoid of outflows including the filaments that are aligned along
 the magnetic fields (see Figure~\ref{fig6}). This could also be an indication for the presence of both the mechanisms that
 lead to parallel and perpendicular outflows.


\section{Conclusions}
\label{sec:summary}
In this comprehensive study, we have investigated the protostellar outflows in 11 massive protoclusters using CO(3--2) line data observed
 with the ALMA. The main results of this study are the following.
 
$\bullet$ We identified a total of 105 outflow lobes in these 11 protoclusters, among which 64 lobes are bipolar, and the remaining 41 are
 unipolar in nature. Except for five outflow lobes, the remaining outflow lobes are identified with ALMA 0.9 mm continuum cores (detailed
 results of cores are not presented in this paper).
 
$\bullet$ Statistically the identified outflow lobes have wide range of velocity (10--150 km s$^{-1}$) and plane of sky extents
 (0.1--0.8 pc) with mostly having velocities below 50 km s$^{-1}$ and average projected plane of sky extents of $\sim$0.2 pc.

$\bullet$ Seven out of our 11 targets are embedded in filaments. Analysis of plane of sky orientations of PA$_\mathrm{lobe}$
 with respect to the filaments (i.e., $\gamma_\mathrm{Fil}$) hosting their driving sources, shows no preferred direction.
 
$\bullet$ We have taken into account the plane of sky projection effect on the observed $\gamma_\mathrm{Fil}$ distribution. Theoretical
 cumulative distribution function was constructed using the projected two-dimensional angles of vectors on three-dimension generated utilizing
 Monte-Carlo simulations. The cumulative distribution function of the observed $\gamma_\mathrm{Fil}$ resembles to
 a random orientation of outflow lobes with respect to the filaments.

$\bullet$ No correlation is also found for the PA$_\mathrm{lobe}$ values with respect to the large-scale magnetic fields or Galactic
 plane position angles. In fact, outflows in filaments aligned along magnetic field PAs also do not show any preferred orientation. Magnetic
 field is reported to be less important at the core scale dynamics of star formation, and our results are consistent with a relatively
 minor role of the magnetic fields.

$\bullet$ Our result is inconsistent with the observational study of \citet{wang11, wang12} and \citet{kong19} for massive star-forming regions.
 They showed perpendicularly aligned outflows with respect to the filaments. However, our targets are associated with \hii regions,
 and hence, are at later evolutionary stage compared to the IRDC studied by \citet{wang11, wang12} and \citet{kong19}. Thus, the outflow axes
 might depend on the age of a star-forming protocluster.
 
Overall, it might be important to explore several other massive protoclusters to get a statistically significant scenario. It is also
 equally important to explore whether the detailed inner structures of the host filaments, e.g., magnetic field, turbulence, etc, have
 a role in explaining such contrasting scenarios.
\acknowledgments We thank the anonymous referee for the critical comments that have helped us improve the scientific content of the paper.
 TB and KW were supported by the National Key Research and Development Program of China (2017YFA0402702, 2019YFA0405100). KW also acknowledges
 support by the National Science Foundation of China (11973013, 11721303), and a starting grant at the Kavli Institute for Astronomy and
 Astrophysics, Peking University (7101502016). TB acknowledges funding from the National Natural Science Foundation of China (NSFC) through
 grant 11633005 and support from the China Postdoctoral Science Foundation through grant 2018M631241. TB also likes to thank the PKU-Tokyo
 Partner fund. We also acknowledge research support from the NSFC through grants U1631102 and 11373010. This work was carried out in part at
 the Jet Propulsion Laboratory, operated for NASA by the California Institute of Technology. DL acknowledges the support funding from the CAS
 International Partnership Program No. 114A11KYSB20160008, and the National Natural Science Foundation of China No. 11725313. This research
 made use of Astropy, a community-developed core Python package for astronomy \citep{astropy18}. LB acknowledges support from CONICYT
 grant Basal AFB-170002. This paper makes use of the following ALMA data: ADS/JAO.ALMA\#2017.1.00545.S. ALMA is a partnership of ESO
 (representing its member states), NSF (USA) and NINS (Japan), together with NRC (Canada), MOST and ASIAA (Taiwan), and KASI (Republic
 of Korea), in cooperation with the Republic of Chile. The Joint ALMA Observatory is operated by ESO, AUI/NRAO and NAOJ.
\appendix
\section{A. Cloud and Filaments identifications}
\label{appendix}
\begin{figure*}[h!]
\epsscale{0.7}
\plotone{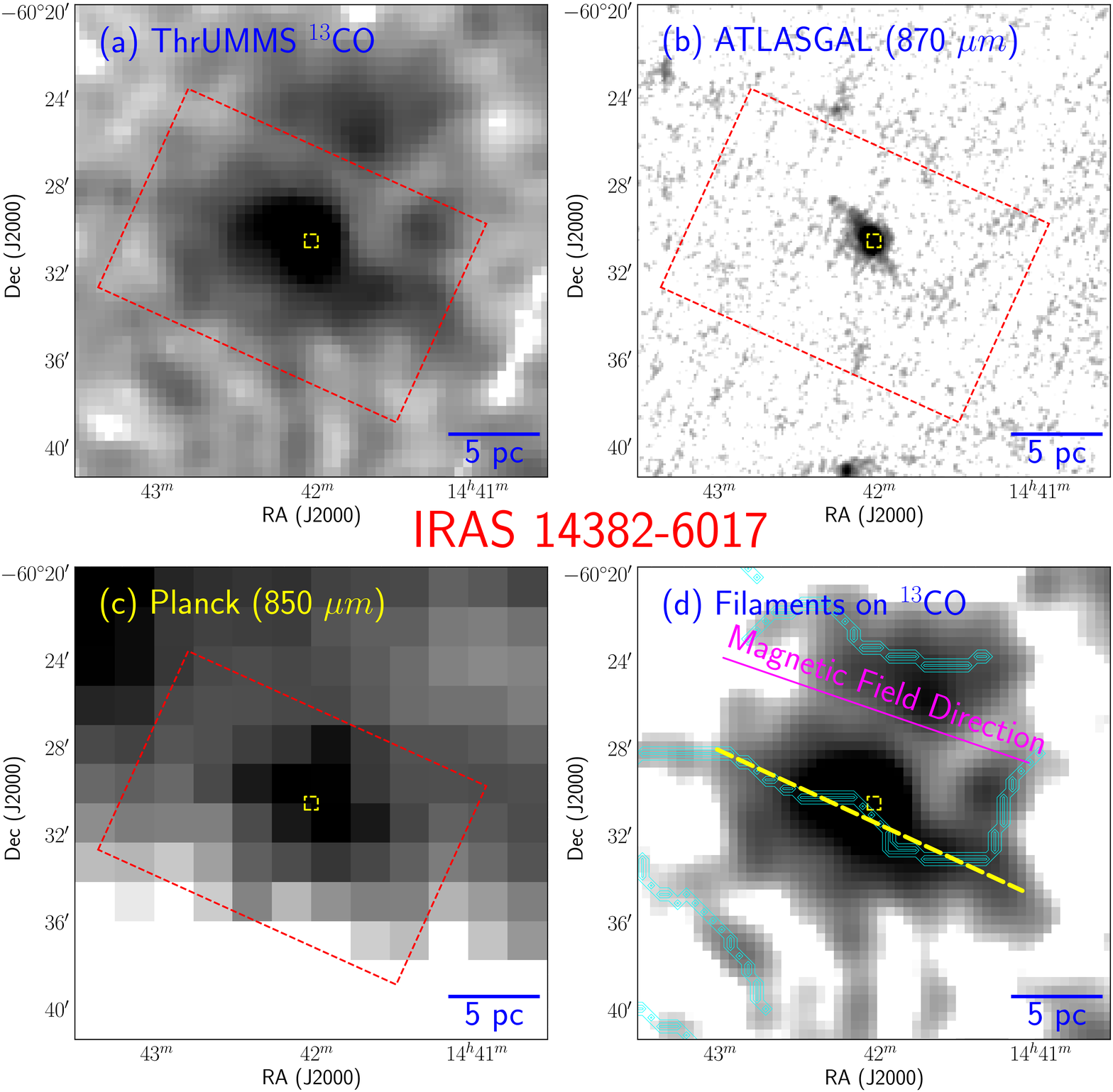}
\caption{Distribution of gas and dust in the IRAS 14382-6017 region. Filament is detected in the integrated $^{13}$CO
 and ATLASGAL 870 $\mu m$ images. Symbols are the same as in Figure~\ref{fig3}. }
\label{figA1}
\end{figure*}

As presented in Section~\ref{SecIdenFil}, we identified the host clouds in the integrated ThrUMMS $^{13}$CO maps and the
 ATLASGAL dust continuum maps. Python-based {\sc filfinder} algorithm \citep{koch15} was also applied on all the identified
 molecular clouds to trace the filamentary structures. Two example figures are presented in Figures~\ref{fig3} and \ref{fig3a}.
 Figures corresponding to the rest of the regions are presented here in two sets: sources with filaments are presented in
 Figure~\ref{figA1}, and sources without filaments are presented in Figure~\ref{figA2}. Note that $^{13}$CO data were not
 available for IRAS 17204-3636 and IRAS 17220-3609 regions. Thus, host clouds in these two regions were identified based on the
 870 $\mu$ ATLASGAL image. 
 
\addtocounter{figure}{-1}
\renewcommand{\thefigure}{\arabic{figure} (Cont.)}
\begin{figure*}
\epsscale{0.7}
\plotone{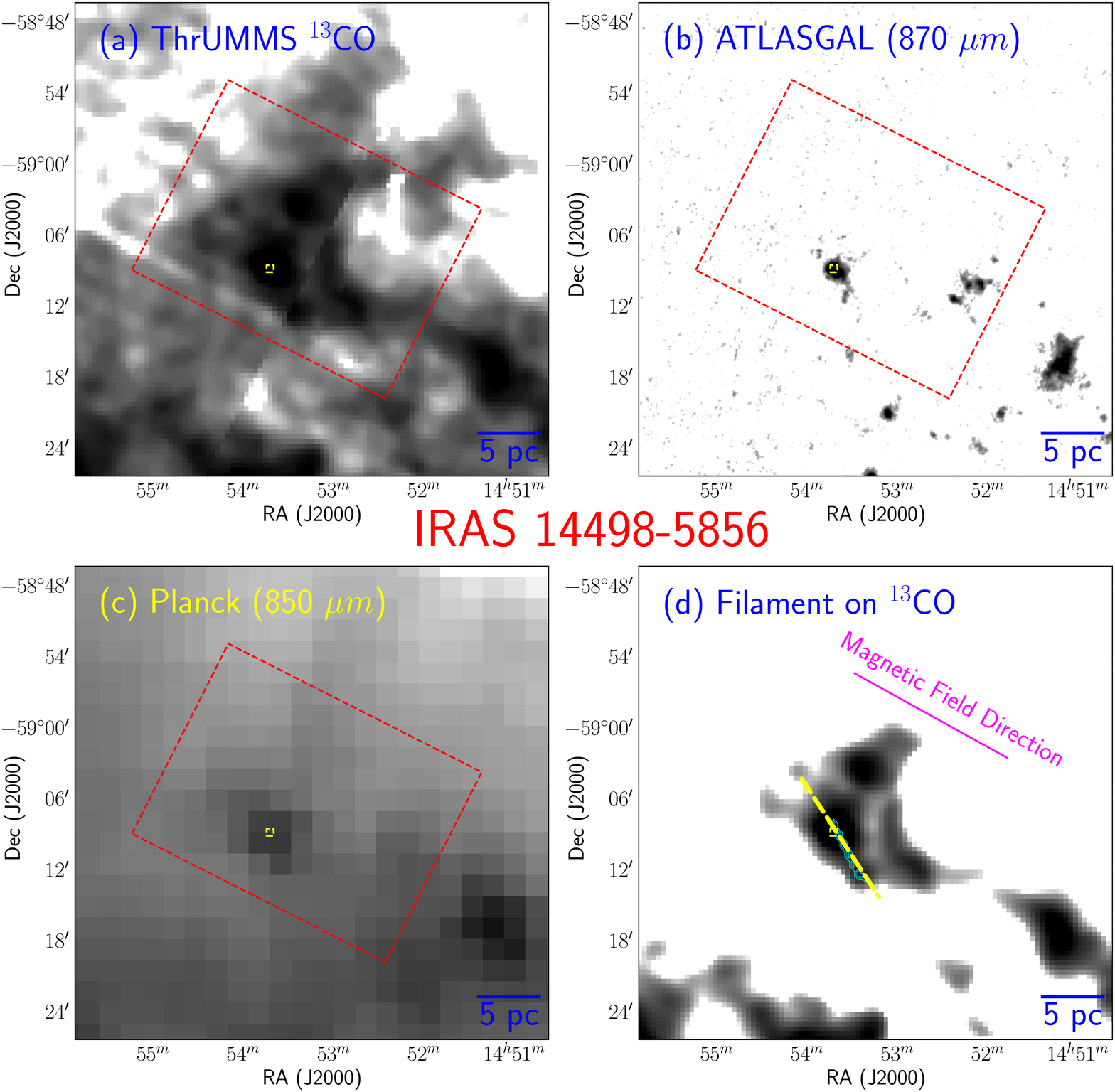}
\caption{Distribution of gas and dust in the IRAS 14498-5856 region. Symbols are the same as in Figure~\ref{fig3}.}
\end{figure*}

\addtocounter{figure}{-1}
\renewcommand{\thefigure}{\arabic{figure} (Cont.)}
\begin{figure*}
\epsscale{0.7}
\plotone{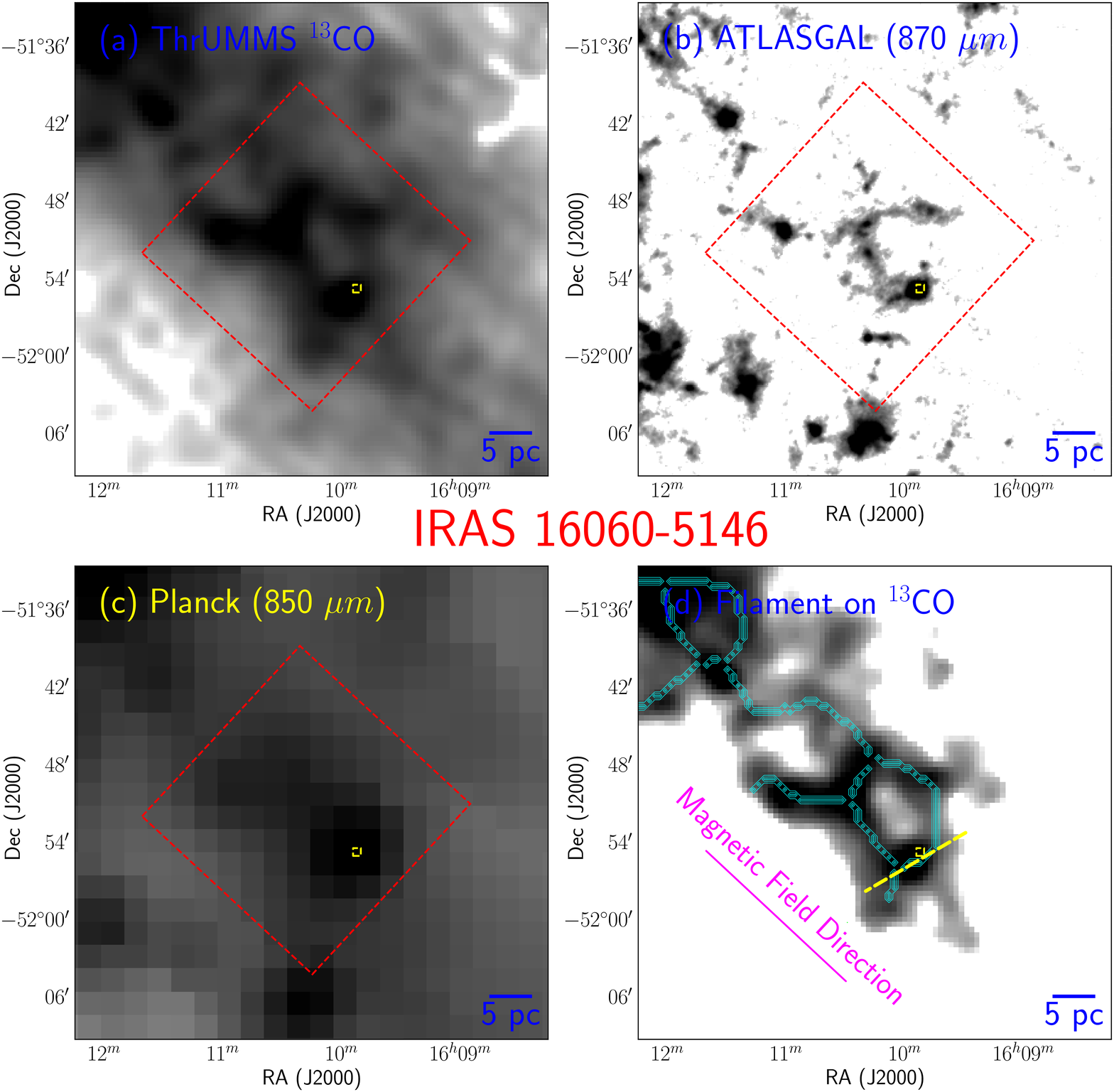}
\caption{Distribution of gas and dust in the IRAS 16060-5146 region. Symbols are the same as in Figure~\ref{fig3}.}
\end{figure*}

\addtocounter{figure}{-1}
\renewcommand{\thefigure}{\arabic{figure} (Cont.)}
\begin{figure*}
\epsscale{0.7}
\plotone{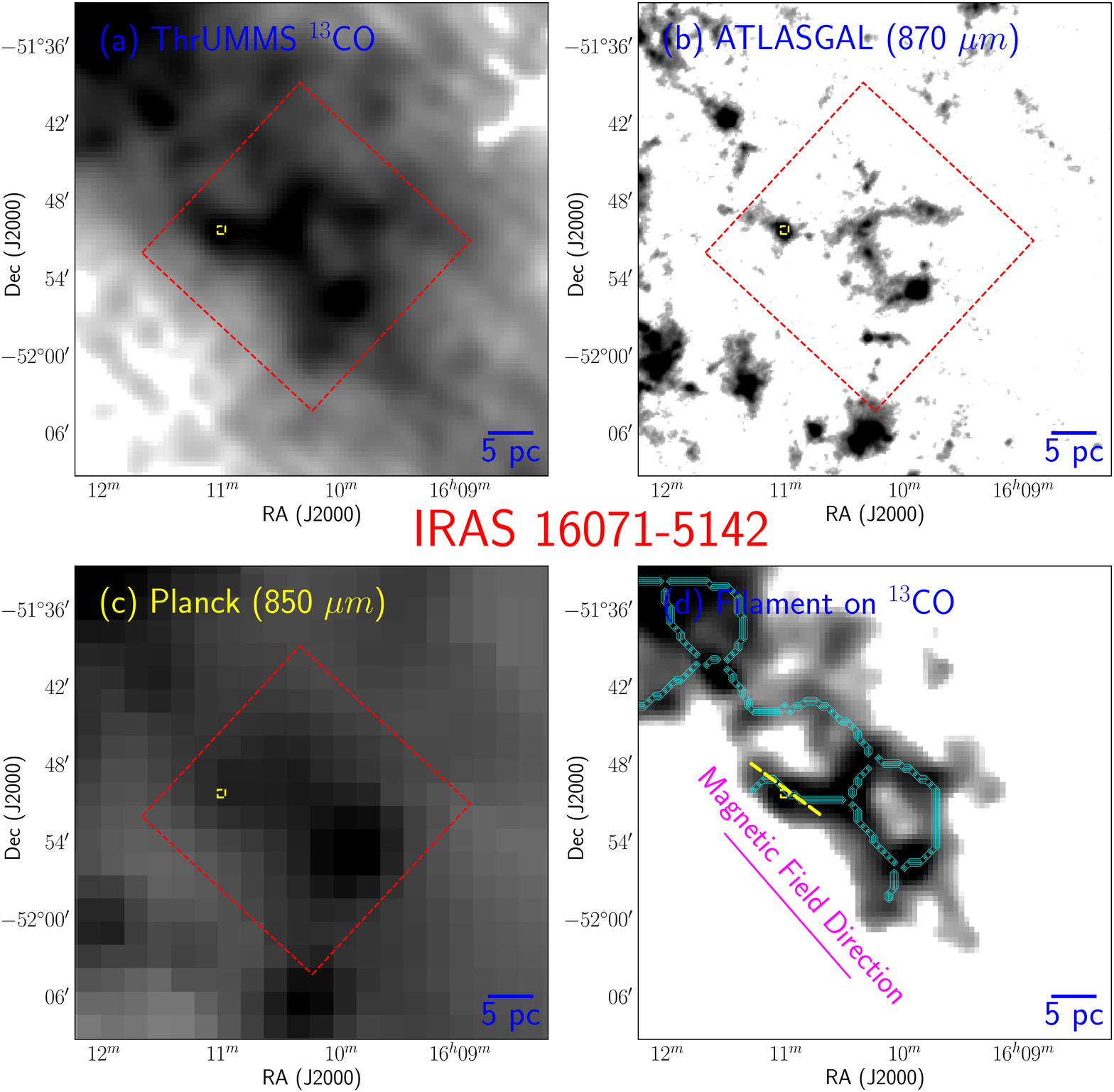}
\caption{Distribution of gas and dust in the IRAS 16071-5142 region. Symbols are the same as in Figure~\ref{fig3}.}
\end{figure*}

\addtocounter{figure}{-1}
\renewcommand{\thefigure}{\arabic{figure} (Cont.)}
\begin{figure*}
\epsscale{0.7}
\plotone{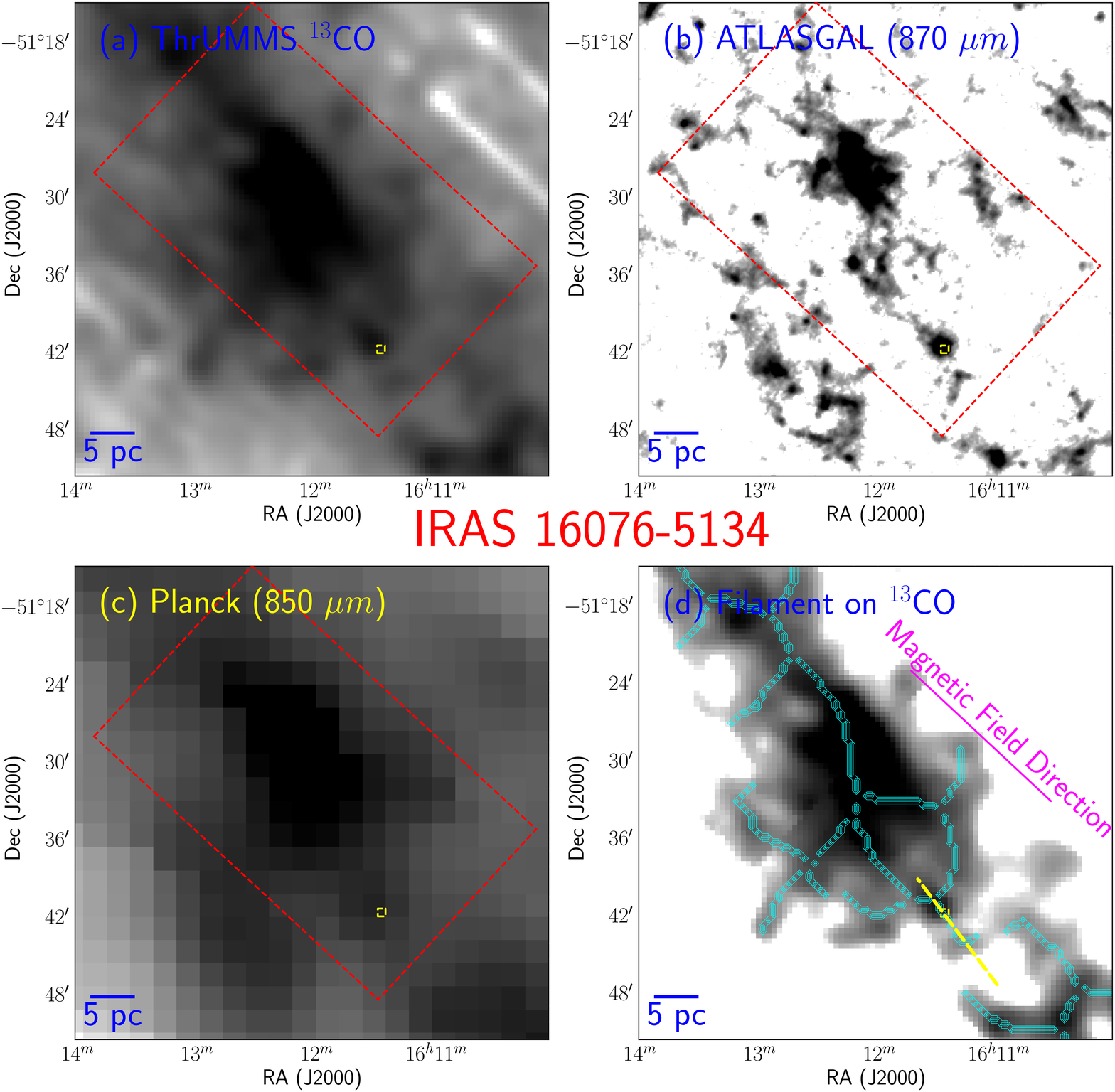}
\caption{Distribution of gas and dust in the IRAS 16076-5134 region. Symbols are the same as in Figure~\ref{fig3}.}
\end{figure*}

\addtocounter{figure}{-1}
\renewcommand{\thefigure}{\arabic{figure} (Cont.)}
\begin{figure*}
\epsscale{0.7}
\plotone{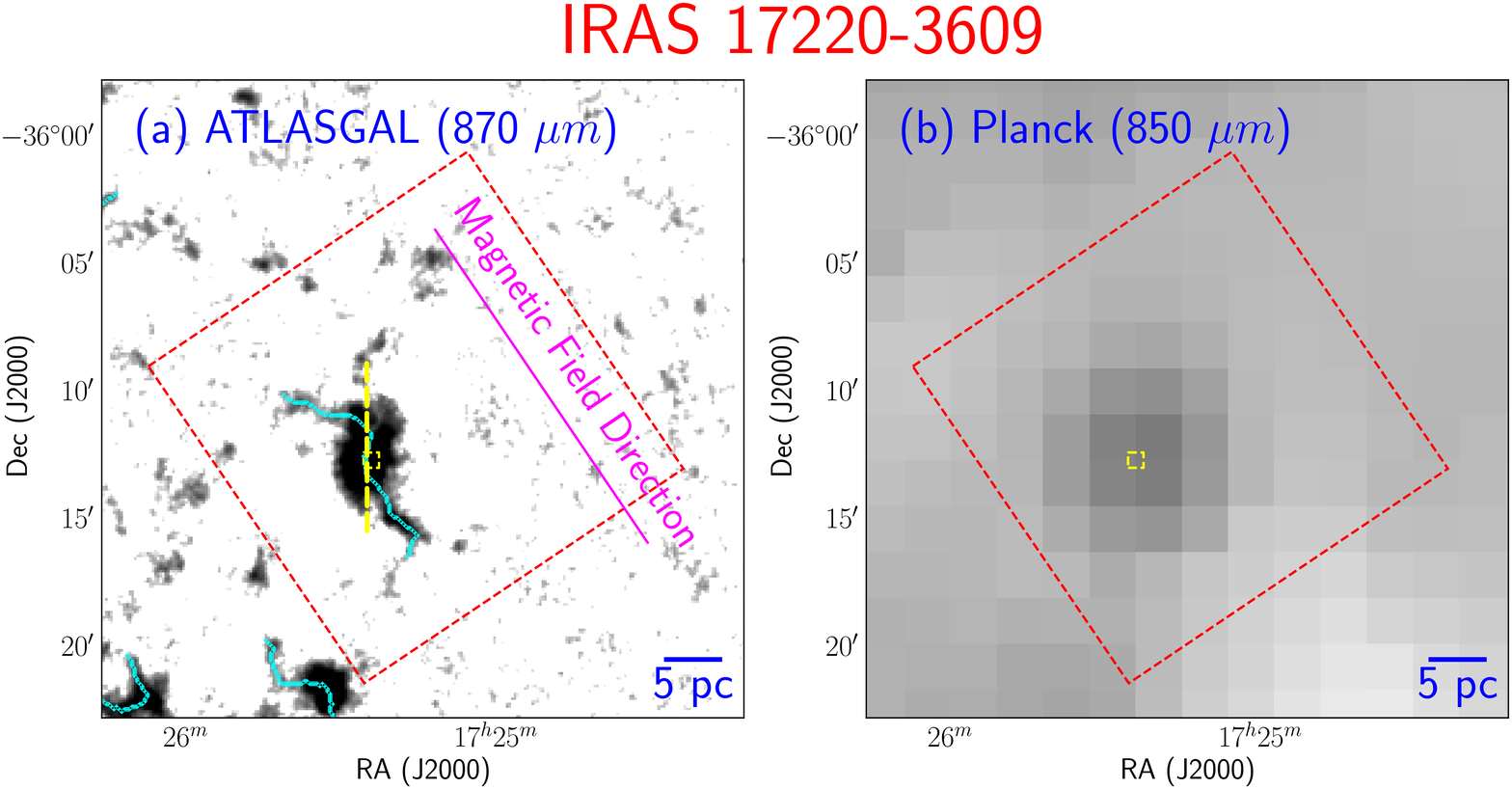}
\caption{Distribution of dust in the IRAS 17220-3609 region. Symbols are the same as in Figure~\ref{fig3}.}
\end{figure*}

\begin{figure*}
\epsscale{0.7}
\plotone{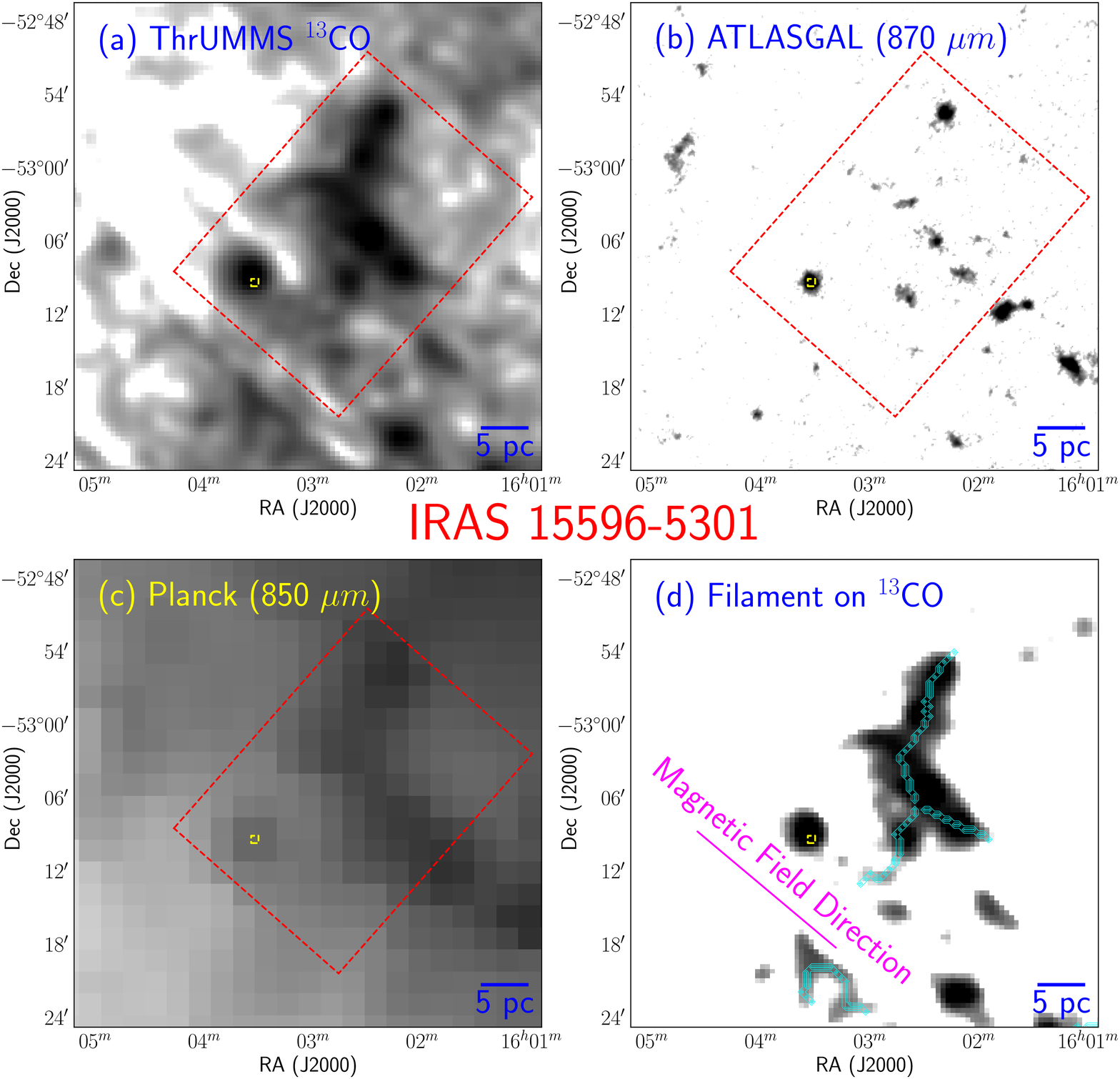}
\caption{Distribution of gas and dust in the IRAS 15596-5301 region where no filamentary structure is detected in the integrated $^{13}$CO
 and ATLASGAL 870 $\mu m$ images. Symbols are the same as in Figure~\ref{fig3}.}
\label{figA2}
\end{figure*}

\addtocounter{figure}{-1}
\renewcommand{\thefigure}{\arabic{figure} (Cont.)}
\begin{figure*}
\epsscale{0.7}
\plotone{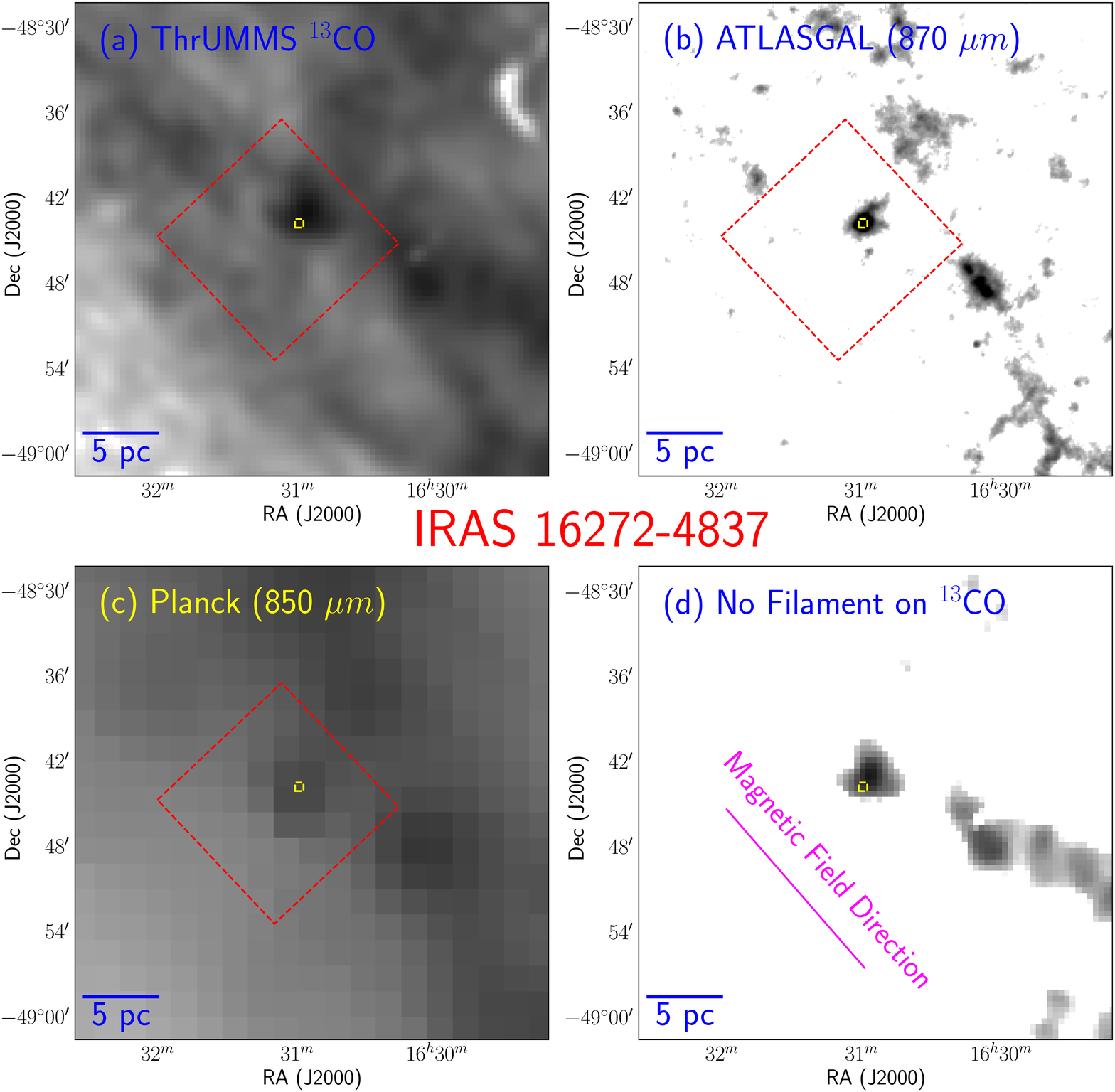}
\caption{Distribution of gas and dust in the IRAS 16272-4837 region. Symbols are the same as in Figure~\ref{fig3}.}
\end{figure*}

\addtocounter{figure}{-1}
\renewcommand{\thefigure}{\arabic{figure} (Cont.)}
\begin{figure*}
\epsscale{0.7}
\plotone{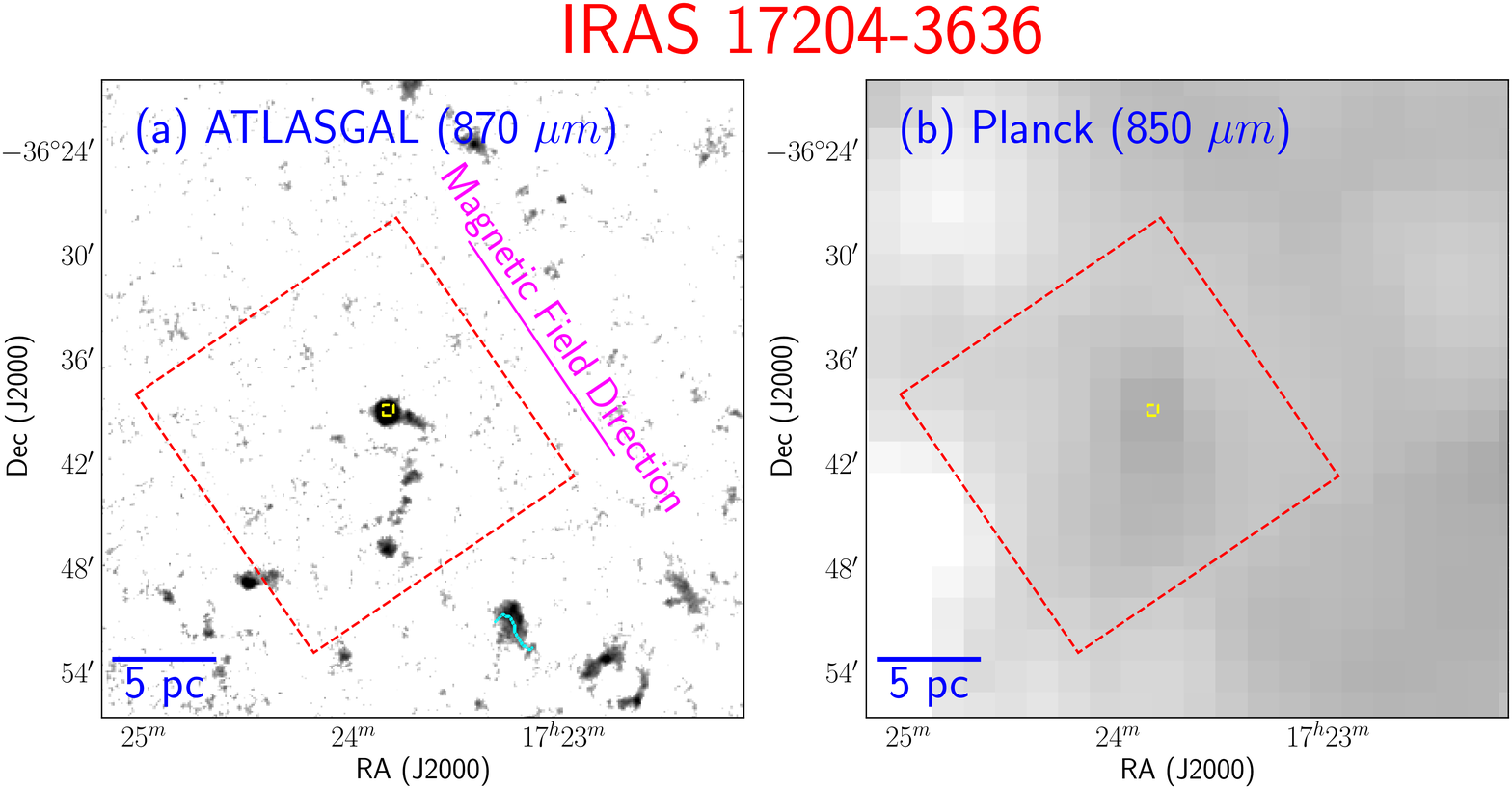}
\caption{Distribution of dust in the IRAS 17204-3636 region Symbols are the same as in Figure~\ref{fig3}.}
\end{figure*}

\section{B. Cumulative distribution of $\gamma_\mathrm{Fil}$}
\label{appendixB}
\begin{figure*}
\epsscale{1.0}
\plottwo{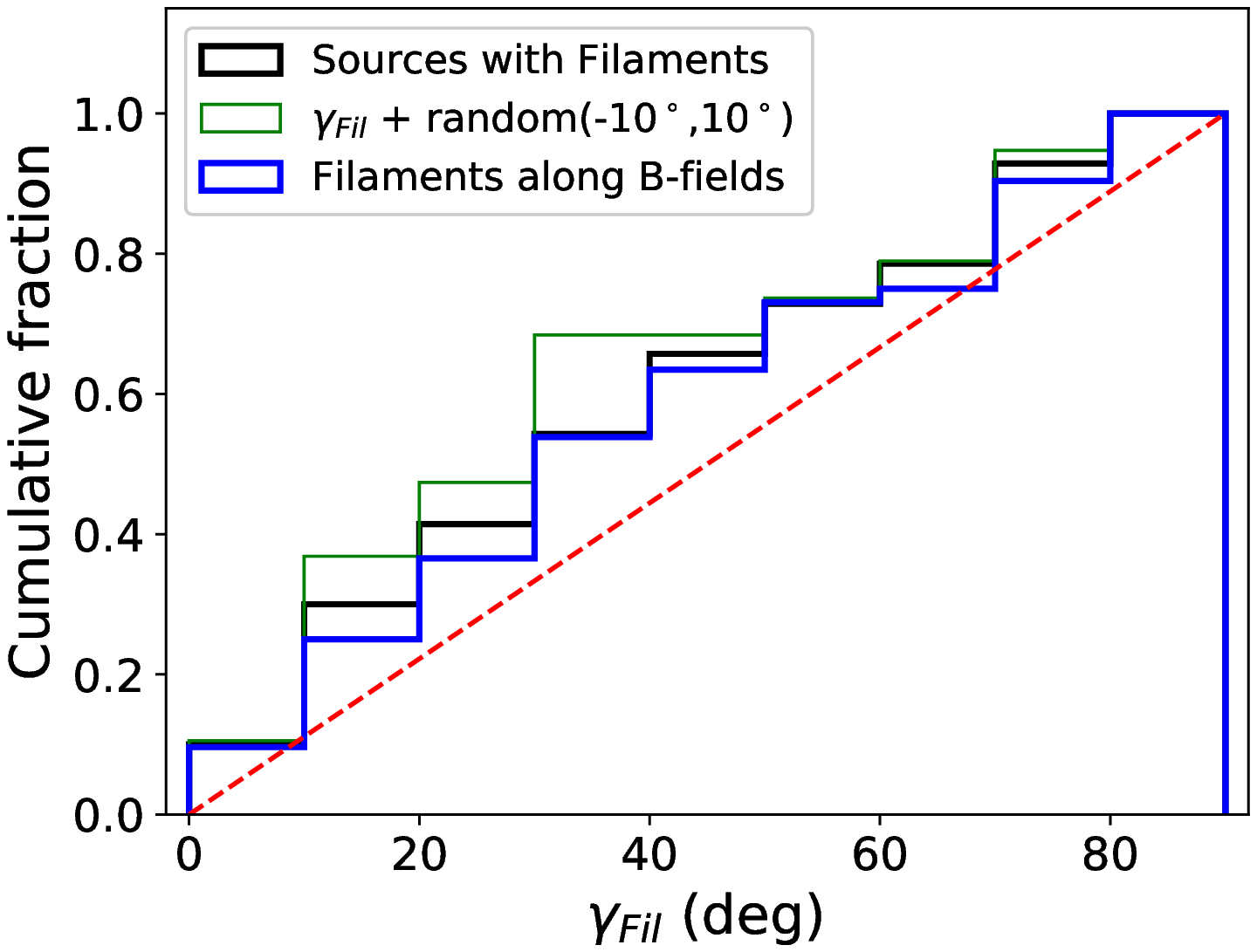}{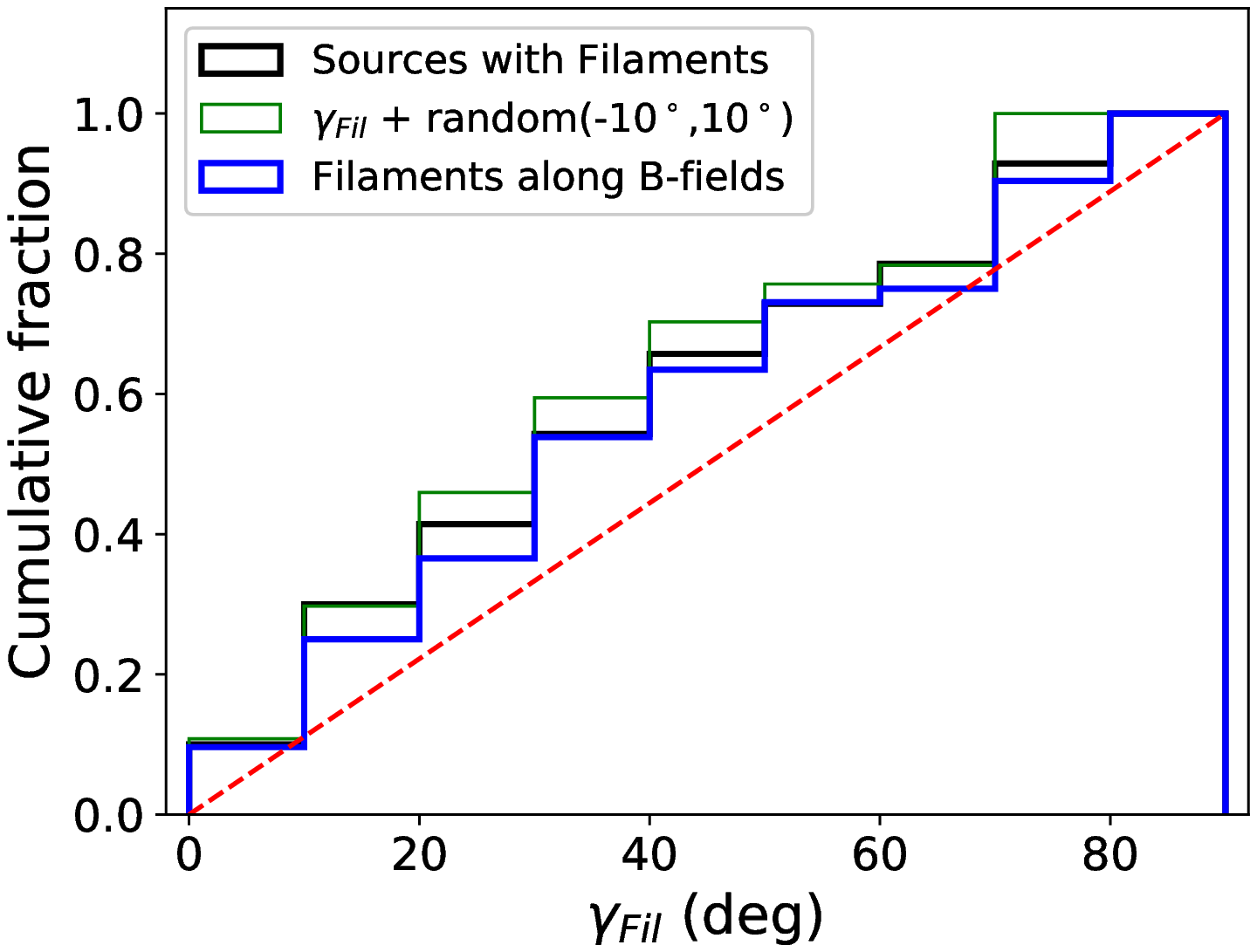}
\caption{Two examples of cumulative distribution functions of $\gamma_{Fil}$ with randomly added value within
 $\pm$10$\degr$. No significant change is noted in the distribution.}
\label{figA10}
\end{figure*}

The cumulative distribution of $\gamma_\mathrm{Fil}$ (Figure~\ref{figA10}) after adding a random value within the
 uncertainty limit of $\pm$10$\degr$.

\end{document}